\documentclass{aa}
\usepackage{graphicx}
\usepackage{txfonts}
\usepackage{lscape}
\usepackage{longtable}
\usepackage[switch]{lineno}

\usepackage{float}
\usepackage[allcolors=blue]{hyperref}

\begin{document}

   \title{Extragalactic Stellar Tidal Streams in the Dark Energy Survey}

   \author{Juan Mir\'{o}-Carretero \inst{1,4}, David Mart\' \i nez-Delgado\inst{2}, Maria A. G\'{o}mez-Flechoso \inst{1,3}, Andrew Cooper \inst{5,6,}, Mohammad Akhlaghi \inst{7}, Giuseppe Donatiello \inst{8}, Konrad Kuijken \inst{4}, Dmitry Makarov \inst{9}, Seppo Laine \inst{10}, Santi Roca-F\`abrega \inst{11}
   }

    \institute{Departamento de F{\'\i}sica de la Tierra y Astrof{\'\i}sica, Universidad Complutense de Madrid, Plaza de las Ciencias 2, E-28040 Madrid, Spain  
    \and
    Instituto de Astrof\'isica de Andaluc\'ia, CSIC, Glorieta de la Astronom\'\i a, E-18080, Granada, Spain 
    \and
    Instituto de F\'isica de Part\'iculas y del Cosmos (IPARCOS), Fac. CC. F\'isicas, Universidad Complutense de Madrid, Plaza de las Ciencias, 1, E-28040 Madrid, Spain
    \and
    Leiden Observatory, Leiden University, P.O. Box 9513, 2300 RA Leiden, The Netherlands
    \and
    Institute of Astronomy and Department of Physics, National Tsing Hua University, Kuang Fu Rd. Sec. 2, Hsinchu 30013, Taiwan
    \and
    Center for Informatics and Computation in Astronomy, National Tsing Hua University, Kuang Fu Rd. Sec. 2, Hsinchu 30013, Taiwan
    \and
    Centro de Estudios de F\'isica del Cosmos de Arag\'on (CEFCA), Unidad Asociada al CSIC, Plaza San Juan 1, 44001 Teruel, Spain
    \and
    UAI - Unione Astrofili Italiani /P.I. Sezione Nazionale di Ricerca Profondo Cielo, 72024 Oria, Italy
    \and
    Special Astrophysical Observatory of the Russian Academy of Sciences, Nizhnij Arkhyz, 369167, Russia
    \and
    IPAC, Mail Code 314-6, Caltech, 1200 E. California Blvd., Pasadena, CA 91125, USA
    \and
    Lund Observatory, Division of Astrophysics, Department of Physics, Lund University, Box 43, SE-221 00 Lund, Sweden
}
\titlerunning{Extragalactic Stellar Tidal Streams in the Dark Energy Survey}
\authorrunning{Mir\'{o}-Carretero et al.}

 
  \abstract
   {Stellar tidal streams are a key tracer of galaxy evolution and have the potential to provide an indirect means for tracing dark matter. For the Local Group, many diffuse substructures have been identified and their link to galaxy evolution has been traced. However, the Local Group does not offer a statistically significant sample of stellar tidal streams. Thus, an analysis of a larger sample beyond the Local Group is required to better probe the frequency and characteristics of these streams to verify whether these properties are in agreement with the predictions of the $\Lambda$CDM model and its implementation in cosmological simulations, taking into account the impact of the baryonic physics modelling.}
   {The main scope of the {\it Stellar Stream Legacy Survey} is to obtain a statistically significant sample of stellar streams in the Local Universe to be able to trace and study minor mergers and their contribution to galaxy evolution with respect to the $\Lambda$CDM theory. For that purpose, we are carrying out the first systematic survey of faint stellar debris from tidally disrupted dwarf satellites around nearby galaxies up to a distance of 100 Mpc.}
   {In this paper, we present a catalogue with the results of the first harvest of stellar tidal streams found by visual inspection in deep images of $\sim$ 700 galaxies from the Dark Energy Survey (DES). We also include, for the first time, a photometric characterisation of the streams obtained by measuring their surface brightnesses and colours.}
   {We found a total of 63 streams in our sample at distances between 40 and 100 Mpc, including 59 which were not previously reported. We measured their average surface brightness for the $g$-band, the $r$-band and the $z$-band, to be 28.35$\pm$0.20,  27.81$\pm$0.13 and 27.62$\pm$0.09 mag arcsec$^{-2}$, respectively. By applying a statistical analysis to our findings, we obtained a stream detection frequency of  9.1\% $\pm$ 1.1\%  for the given surface brightness limit of the DES image sample, in agreement with previous studies. We identified stream progenitors in 5--14\% of our stream sample, depending on the confidence level.}
  {The first catalogue of streams in the Local Universe presented here will be complemented by future stream surveys within the {\it Stellar Stream Legacy Survey} and can be exploited in studies pertaining to galaxy evolution and cosmological models. In this work we have learnt that the faintest measured stream surface brightness can be significantly brighter than the surface brightness limit of an image measured at pixel level (in our case up to $\sim$ 1 mag arcsec$^{-2}$ for the $r$ band) mainly due to correlated noise present in the images.}

   \maketitle
%


\section{Introduction}
\label{sec:introduction}

The hierarchical framework for galaxy formation and evolution predicts that the evolution of massive galaxies is influenced by mergers with low-mass satellite galaxies. The stellar halos of galaxies are expected to be rich in remnants from such events. Stellar streams are the remnants of so-called minor mergers, mergers in which the difference in mass of the two galaxies is significant. A mass ratio lower than $1/3$ is sometimes quoted in the literature for minor mergers, see \citet{newberg2016}, with other authors have considered a limit ratio of $1/10$, see \citet{mancillas2019}. 
Tidal streams are 
a particular case of low surface brightness (LSB) structures and 
not to be 
confused with tidal tails; 
a morphology type 
that can also be the products of 
major mergers, mergers between galaxies with a mass ratio greater than $1/3$ \citep{toomre1972}. It has been suggested, on the basis of observations and simulations, that minor mergers are more frequent than major mergers in the nearby Universe \citep{guo2008,jackson2022}.

By studying stellar streams, in addition to learning about their evolution history of their host galaxies, other key cosmological questions can also be addressed. By comparing observations and simulations, we can verify whether the frequency and characteristics of streams in the Local Universe are consistent with predictions based on the $\Lambda$CDM paradigm, thus providing clues to better modelling of the baryonic processes within dark matter halos \citep{shipp2023}.   

Much research has been carried out on stellar tidal streams, however mostly limited to streams in the Milky Way and to a lesser extent in the Local Volume \citep{belokurov2006,martinez-delgado2010,hood2018,shipp2018,ferguson2022,li2022}. 
With time, detection methods have improved to facilitate the detection of streams with the support of 
algorithms to increase automation; for example, \citet{ibata2019} and \citet{ibata2021a} use STREAMFINDER in combination with astrometric data from Gaia to detect streams in the Milky Way. \citet{kado-fong2018} present an algorithm to detect tidal features in HSC-SSP data that delivers 
50$\%$ completeness for tidal features with a surface brightness of $\sim$  26.4 mag arcsec$^{-2}$ .

Earlier work has been based on the exploitation of first generation wide-field, digital imaging surveys, most notably SDSS \citep{york2000}. However, the search for stellar tidal streams beyond the Local Volume 
with these surveys is severely hampered by the difficulty of detecting faint structures at their typical surface brightness limit of $\sim$ 26 mag arcsec$^{-2}$ . Valuable efforts have been made to detect streams beyond the Local Volume, but their results have been somewhat limited by the insufficient depth of the images \citep[e.g.][]{miskolczi2011}. 
Recently, images from deeper, wide-field imaging surveys have been made available, 
including the DESI Legacy Surveys \citep{dey2019}, which provide 
the opportunity to advance the study of stellar tidal streams in the Local Universe.

Relevant work to characterise tidal features beyond the Local Volume has recently been carried out by \citet{sola2022}, who have characterised the morphology of more than 350 low surface brightness structures up to a distance of 42 Mpc through annotation of images from the Canada--France Imaging Survey (CFIS2) and the Mass Assembly of early-Type gaLAxies with their fine Structures survey (MATLAS3) by a team of experts using a graphical tool. The morphology of the remnants of galaxy mergers is relevant as it can provide information about the type and the time of the merger. In \citet{giri2023}, remnants of recent mergers in nearby early-type galaxies are studied and their histories inferred from the morphology of tidal features identified through observations, but this work, as
earlier work by \citet{tal2009}, focuses on early type (elliptical) galaxies only.

Besides the depth of the images, a further difficulty in detecting stellar tidal streams is that 
standard pipelines for reducing observational data are generally geared to the production of catalogues for point sources, and tend to over-subtract the image sky background, rendering the detection of faint sources difficult or even impossible. With the advent of modern detection tools developed with faint structures in mind, the opportunity to get more faint source detection out of the available survey images has increased, see \citet{haigh2021} and \citet{kelvin2023} for a comparison of these tools.

 The main observational objective of the {\it Stellar Stream Legacy Survey (SSLS)} is to successfully complete a systematic survey of stellar tidal streams in a parent galaxy sample of  $\sim$ 3200 nearby galaxies (which includes $\sim$ 800 Milky Way analog systems). For this purpose, we have generated a catalog of deep, wide-field images of each target galaxy by co-adding existing calibrated data from the recently completed DESI Legacy Survey imaging surveys, as described in \citet{martinez-delgado2023b} (referred to as \textit{Paper1} in the rest of this article). Assuming observable streams are present in 
 approximately 10--15$\%$ of sample galaxies for our target depth \citep[e.g.][]{morales2018,miro-carretero2023}, we expect to compile a sample of up to $\sim400$ galaxies with one or more streams. This will allow us to construct, for the first time, robust distributions of key properties for a statistically significant sample of stellar tidal streams in the Local Universe, including their surface brightness, colour, width, length and distance to their host, alongside the same properties for their possible progenitors. 
 For the type of mergers addressed in our work, namely minor mergers, the morphology of the resulting streams can give us information about the orbit and the mass of the progenitor, as well as about the time when the accretion began.
 This extensive database will provide observational constraints on the frequency, luminosity/mass distribution, size and incidence of morphological types of these tidal structures in the Local Universe, which can directly be compared with results from cosmological simulations. 

In this paper, we present the first catalog and photometry of stellar streams found in the Dark Energy Survey footprint, one of three extensive sky areas contained in our SSLS (see Figure 1 in \textit{Paper1}.
In Section \ref{sec:methodology} we introduce the image sample used as input for this work, along with the corresponding selection criteria, and the methods and tools applied for measuring the photometry parameters in those images. In Section \ref{sec:catalogue} we  present the content of the stellar stream legacy survey catalogue for the DES images. Specific aspects of the results, such as frequency, morphology etc., as well as statistical considerations, are discussed in Section \ref{sec:results}. The conclusions and future work are summarised in Section \ref{sec:discussion}. 

\begin{figure*}
\centering
  \includegraphics[width=1.0 \textwidth]{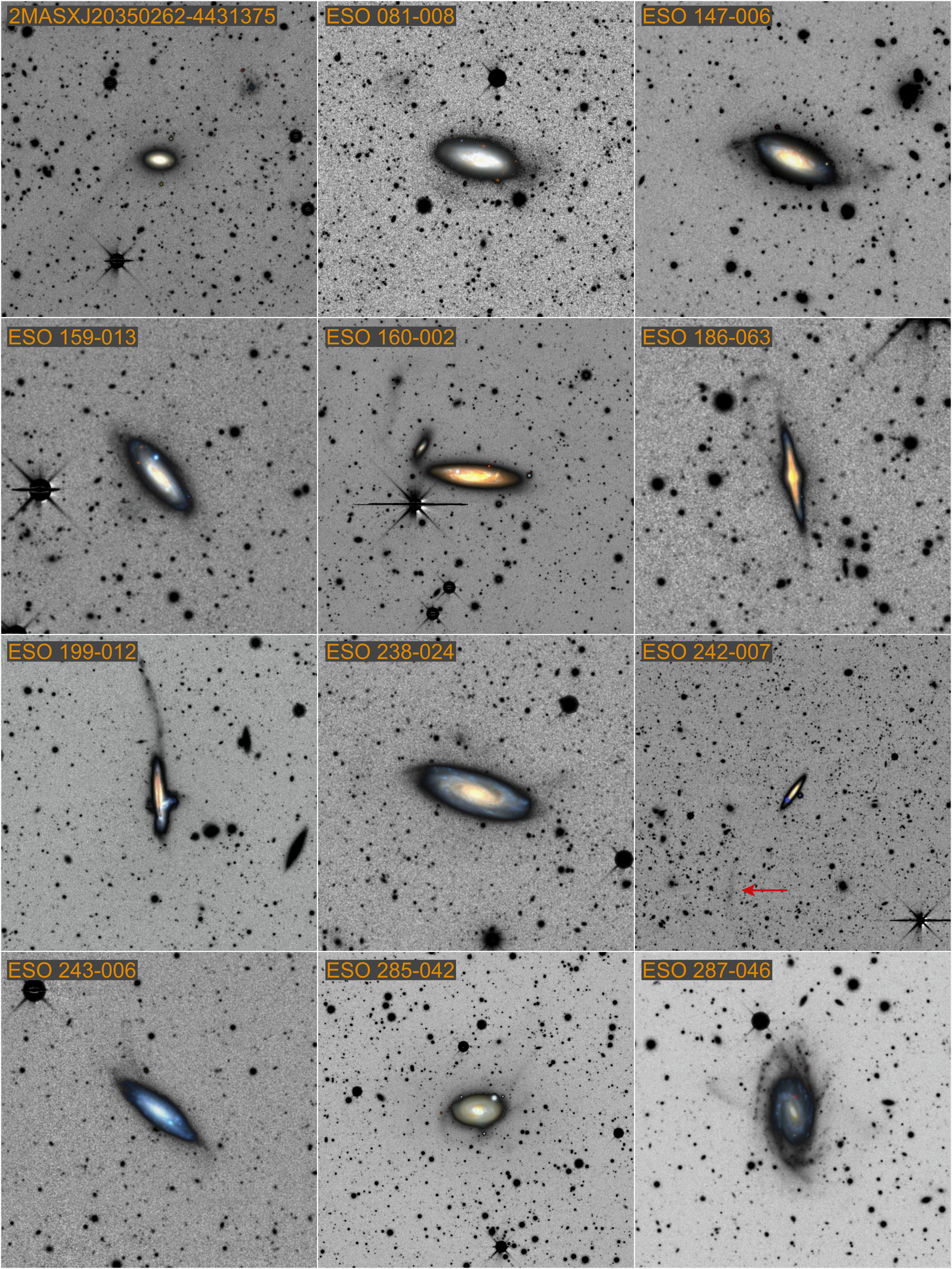}
  \caption{\textbf{A)} Sample of DES images used as input to the analysis. Red arrows point to those (parts of) streams that are difficult to see in the images.}
  \label{fig-sample1}
\end{figure*}

\addtocounter{figure}{-1}

\begin{figure*}
\centering
  \includegraphics[width=1.0 \textwidth]{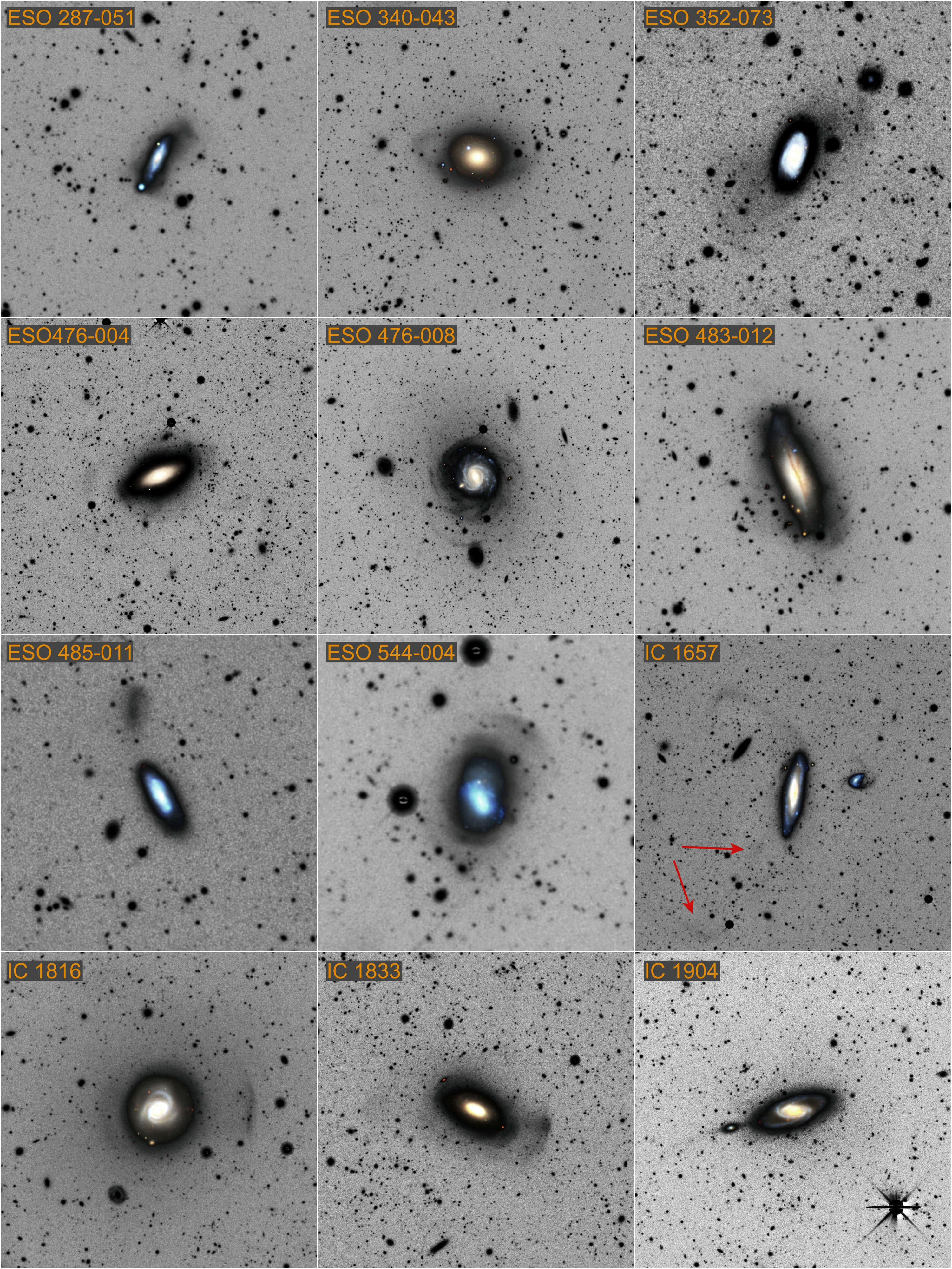}
  \caption{\textbf{B)} Sample of DES images used as input to the analysis. Red arrows point to those (parts of) streams that are difficult to see in the images.}
  \label{fig-sample2}
\end{figure*}

\addtocounter{figure}{-1}

\begin{figure*}
\centering
  \includegraphics[width=1.0 \textwidth]{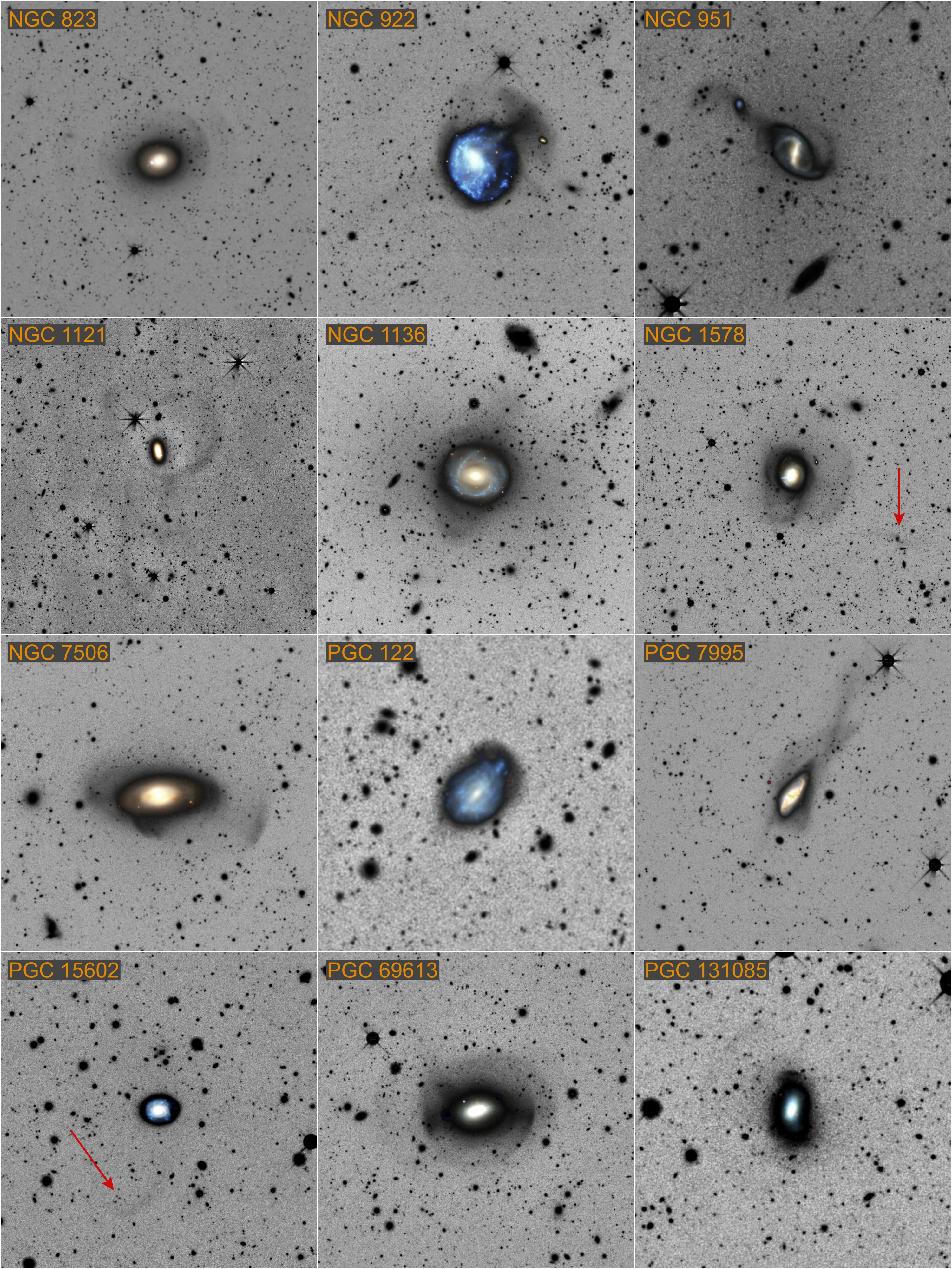}
  \caption{\textbf{C)} Sample of DES images used as input to the analysis. Red arrows point to those (parts of) streams that are difficult to see in the images.}
  \label{fig-sample3}
\end{figure*}

\addtocounter{figure}{-1}

\begin{figure*}
\centering
  \includegraphics[width=1.0 \textwidth]{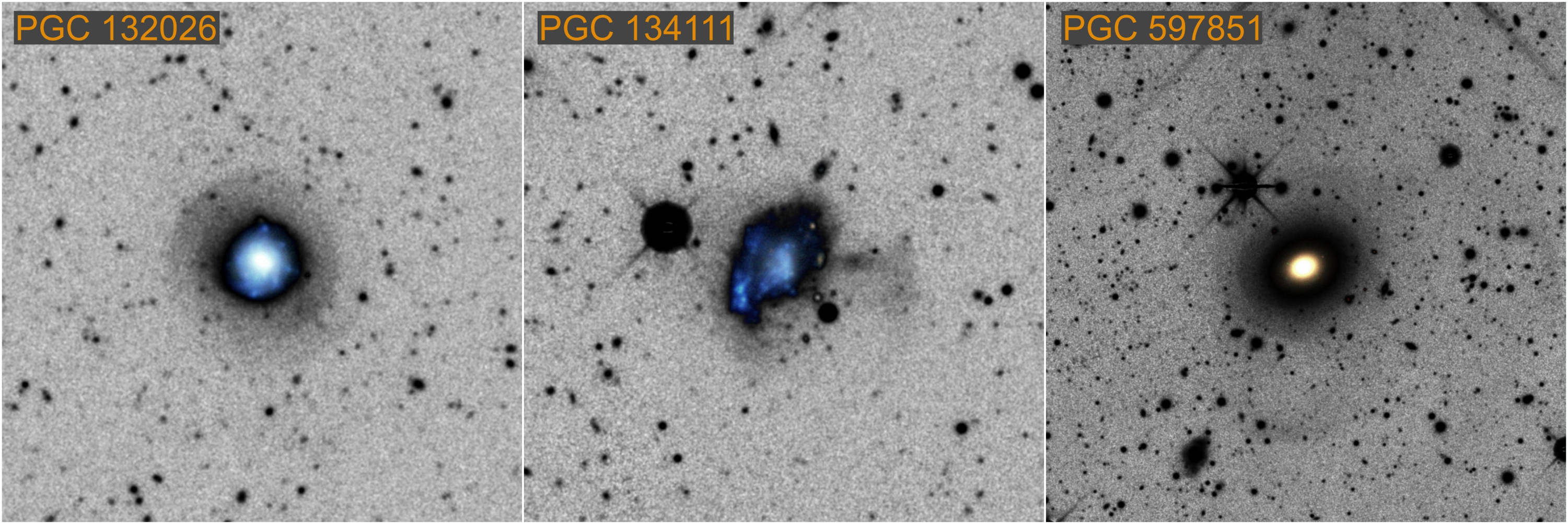}
  \caption{\textbf{D)} Sample of DES images used as input to the analysis.}
  \label{fig-sample4}
\end{figure*}

\section{Methodology}
\label{sec:methodology}

\subsection{ Imaging data and target sample}
\label{sec:imagesample}

Our stream survey is implemented in essentially three steps: 1) To produce deep images for each galaxy from a well-defined parent sample using available imaging survey data; 2) To search systematically for tidal streams by means of visual inspection; and 3) To analyse photometrically the images, verifying the detection and measuring the surface brightness and colours of any streams detected, along with an assessment of detection frequency and limits. The ultimate goal of our approach is to compare these results quantitatively with those of cosmological simulations, by creating a catalog of mock images that include realistic observational artifacts and systematic errors (the results will be published in a follow-on paper). 

For the first step, we exploit public, deep imaging data released recently by a new generation of large scale photometric surveys in three optical bands ($g$,$r$ and $z$), as described in \textit{Paper1} (see their Figure 1). For this paper, we have used the data from the Dark Energy Survey (DES), which  covers 5000 square degree of the southern sky with photometric observations in the grizY bands. It uses the Blanco 4m telescope at the Cerro Tololo Interamerican Observatory in northern Chile, with the 3 square degree Dark Energy Camera (DECam) imager, a 570 Megapixel CCD camera. The observations were carried out between 2013 and 2019 (for more details see https://www.darkenergysurvey.org/).

First, the selection of the complete isolated galaxy sample of the SSLS was made using the criteria described in \textit{Paper1} (see their Section \textit{2.3 Survey design and sample selection}), based on luminosity and recessional velocity (as a proxy for distance) taken from the Hyperleda database \citep{makarov2014}, including a total number of $\sim$ 52600 galaxies with redshift $z < 0.02$. For this paper, we selected only those galaxies whose position is within the DES footprint. The precise criteria were: i) the absolute K-band magnitude should be < -19.6 mag (note that this is also a proxy for stellar mass), ii) the radial velocity (VLG) in the Local Group rest-frame should be 2000 [km/s] < VLG < 7000 [km/s] (30 [Mpc] < D < 100 [Mpc], D being the distance from the Sun, iii) the galaxy should be outside of the Galactic plane, with |b| > 20 deg and iv) there is no neighbour brighter than $K_{\rm gal}=2.5$~mag with $|\Delta V|<250$ km s$^{-1}$ within a projected radius of 1~Mpc around each target (isolation criterion). 
These distances were calculated assuming a standard flat $\Lambda$CDM cosmology with Hubble parameter $H_\mathrm{0}=73\,\mathrm{km\,s^{-1}\,Mpc^{-1}}$
The resulting DES sample contains 689 galaxies, of which around 90\% are spiral galaxies, in agreement with the well-known galaxy density-morphology relation, given the galaxy isolation criterion applied to our sample selection. 

Subsequently, the images of the resulting sample were visually inspected  using the Legacy Survey Sky Viewer \footnote{\url{https://www.legacysurvey.org/viewer}} and a subset of targets for which stellar tidal streams were suspected by eye were selected for further analysis. New coadded images centered on these selected targets were then computed for this project from the individual exposures of the {\it DESI Legacy Imaging Surveys} \citep[][; LS]{dey2019}, using a modified version of the Legacy Survey reduction pipeline {\it Legacypipe} that modifies the way the image backgrounds (``sky models'') are computed \footnote{\url{https://legacypipe.readthedocs.io/en/latest/}}. By default, {\it Legacypipe} uses a flexible spline sky model which can over-subtract the outskirts of large galaxies \citep{moustakas2023}. The modified version we use instead subtracts the sky background from each individual exposure, using a custom algorithm that preserves the low-surface-brightness galactic features of interest. We first minimized the relative background levels between the overlapping CCD frames in each band, and then, after detecting and masking sources as well as Gaia stars, we subtracted the sigma-clipped median in the outer half of the image cutout (see \textit{Paper1} for details). Examples of the input images are shown in Figures \ref{fig-sample1}\textbf{A)}, \ref{fig-sample2}\textbf{B)}, \ref{fig-sample3}\textbf{C)} and \ref{fig-sample4}\textbf{D)}. We then carried out an analysis of these images (see following section) which allowed us to confirm that the suspected stream was indeed a remnant of a minor merger through the measurement of its morphological and photometric parameters.

\subsection{Photometry analysis}
\label{sec:dataanalysis}


\begin{figure*}
\centering
    \includegraphics[width=1.0 \textwidth]{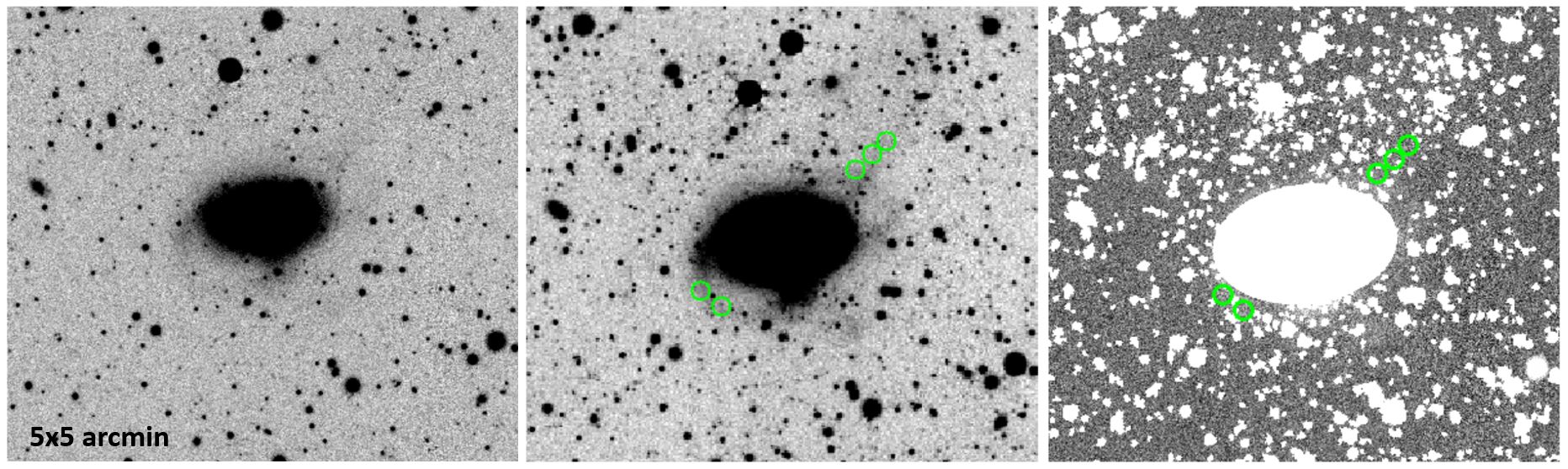}
   \caption{Photometry measurement method with Gnuastro: example of a stream around ESO  285-042. \textit{left:} $r$-band background-subtracted input image; \textit{center:} apertures tracing the stream in which photometry parameters, such as magnitude, surface brightness and colours are measured (image resampled to a  pixel size $4\times4$ larger than the input image); \textit{right:} image with a masked central galaxy, as well as foreground and background sources.}
    \label{fig-method}
\end{figure*}  

\begin{figure}
\centering
    \includegraphics[width=0.8 \columnwidth]{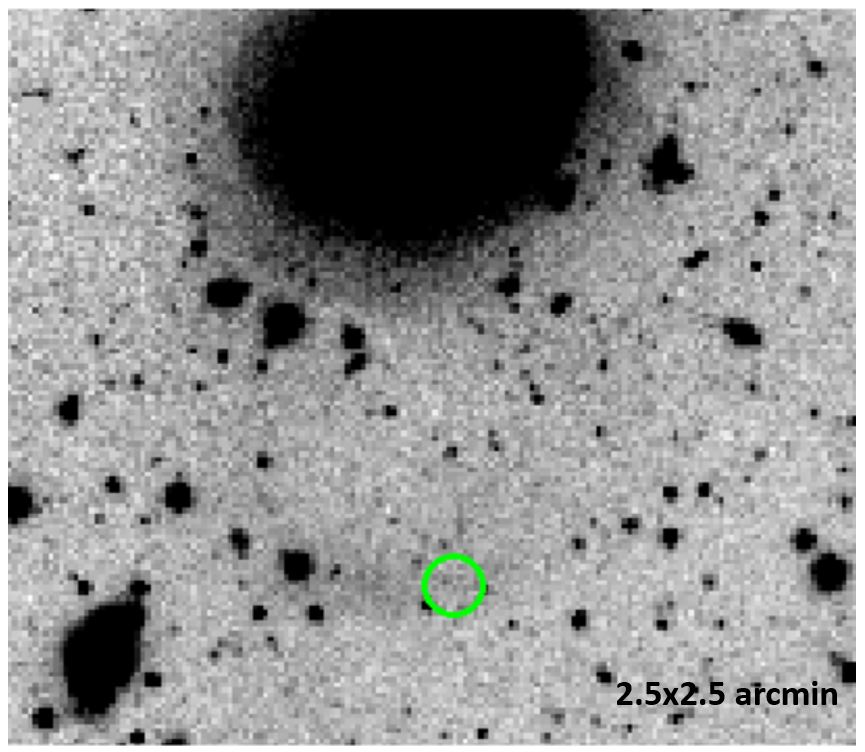}
    \includegraphics[width=0.8 \columnwidth]{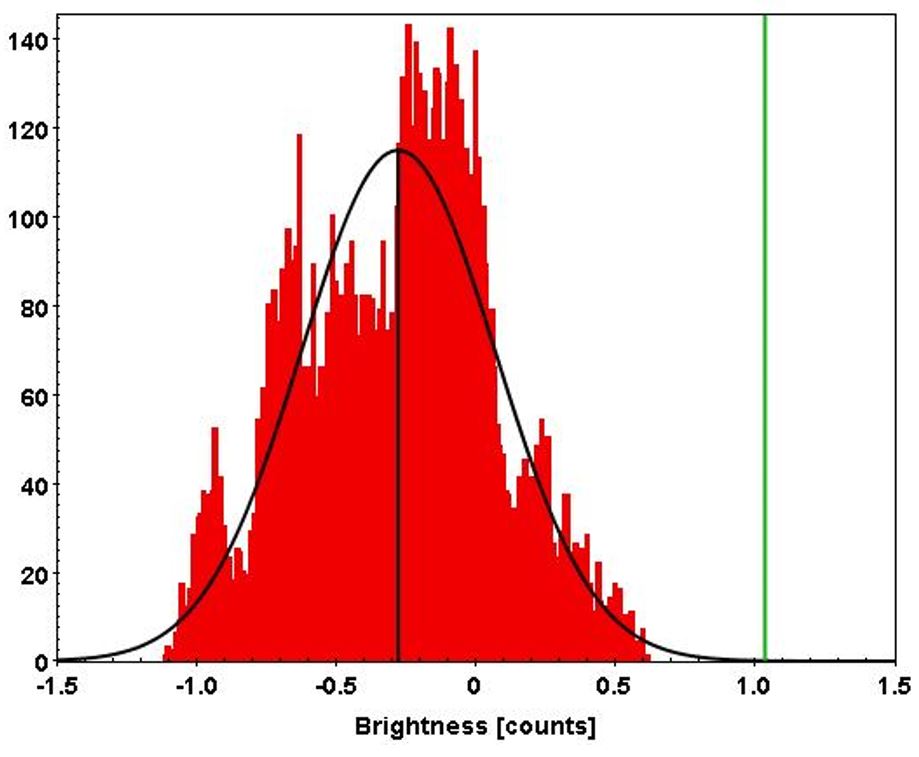}
   \caption{Detection Significance Index (DSI). \textit{top:} image showing an aperture of 74.41 arcsec$^{2}$ (net area excluding pixels identified as background or foreground sources) placed on a faint segment of the stream around PGC 597851; \textit{bottom:}  histogram showing the flux distribution in 10000 apertures of the same size, randomly placed over non-detection regions of the image. The green vertical line shows the flux in the green aperture above, with a measured surface brightness of 26.91 mag arcsec$^{-2}$ , at 3.73 $\sigma$ (DSI) from the distribution median. Note that the histogram is not symmetric, as different regions can have different depths.}
    \label{fig-DSI}
\end{figure}  

Our photometric analysis includes  measurements of the surface brightness in the LS $r$, $g$ and $z$ bands for each stream, as well as for their candidate progenitor satellite galaxy (when identified). Taking advantage of the depth and photometric quality of the LS survey images, and the availability of images in three bands, we also measured the colour of the streams. All the magnitudes in this paper are in the AB system \citep{oke1983} unless otherwise noted.

We carried out the photometric analysis with  the state-of-the-art  {\it GNU Astronomy Utilities} (Gnuastro)\footnote{\url{http://www.gnu.org/software/gnuastro}} software. 
We made all the measurements by applying Gnuastro's {\sc MakeCatalog} subroutine \citep{akhlaghi2019a} on the sky-subtracted images generated by Gnuastro's {\sc NoiseChisel} \citep{akhlaghi2015,akhlaghi2019b}.


We use {\sc NoiseChisel}
for the detection of the streams (and all other sources of the images).
{\sc NoiseChisel} was designed specifically to detect low-surface-brightness structures. {\sc NoiseChisel} also calculates the background sky and subtracts it from the input image.  
The subtracted background sky level is not a constant value over the image; the sky is assumed to be constant only on tiles of a configurable number of pixels (tiles of 40x40 pixels, equivalent to $\sim$ 10x10 arcsec has been used in this work), which form a tessellation of many tiles over the image. In this way, the environment of the stream is taken into consideration for the calculation of the sky background to be subtracted locally. For a complete introduction to the robustness of this method, we refer to the corresponding chapter of the Gnuastro book
\footnote{\url{https://www.gnu.org/software/gnuastro/manual/html_node/Skewness-caused-by-signal-and-its-measurement.html}}. Then, segmentation is carried out by {\sc Gnuastro's} {\sc Segment} package, which labels all the detected sources. The foreground and background sources are identified as clumps and are masked before the photometry measurements are carried out by {\sc MakeCatalog}, another package belonging to {\sc Gnuastro}.

Regarding the modelling and subtraction of the host galaxy halo, this approach has been applied by modelling the host halo with a Sersic profile. However, due to the irregular shape of the spiral host galaxies analysed, this technique was difficult to apply, particularly for hosts that are not face-on, and had the effect of over-subtracting the diffuse area around the host; this negatively impacted  the photometry measurement of the stream. Instead, we actually estimated  the zone of influence for every host by measuring the gradient of the surface brightness in its faint surroundings and masking the host to the point of transition to a flat radial surface brightness profile, making sure the apertures where the stream photometry is measured lie outside of such a zone. This is only relevant for those streams that are close to the outskirts of the host galaxy. Fig. \ref{fig-method} illustrates the process applied to measuring stream photometry parameters such as magnitude, surface brightness and colours.

\begin{figure}
\centering
     \includegraphics[width=0.9\columnwidth]{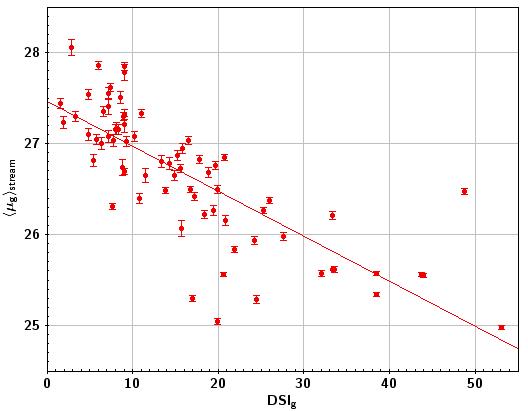}
    \caption{Measured stream average surface brightness (band g) versus average Detection Significance Index.}
    \label{fig-sbvsdi}
\end{figure}

We measured the surface brightness and colours in apertures, placed manually following closely the detection map of the stream generated by {\sc NoiseChisel}, once all foreground and background sources were masked. A succession of circular apertures allows us to measure colour gradients and can easily adapt to the stream contour; however, in a few cases where the stream shape allowed, larger polygonal apertures were used to reduce the measurement uncertainty. The diameter of the circular apertures is as close as possible to the perceived width of the stream. Regions where the stream surface brightness was judged to be significantly blended with light from the host galaxy or that were significantly obscured by clumps were avoided. 
 As an illustration of the method, Figure \ref{fig-method} shows an image of the stream around ESO 285-042 in the $r$-band. The stream photometry parameters such as magnitude, surface brightness and colours are measured in the apertures depicted.
This figure also shows the image with the central galaxy masked out, along with all detected foreground and background sources. We obtained a representative surface brightness and colour for each stream by taking the mean of the individual aperture measurements.

Within each (circular or polygonal) aperture, the flux is measured over every pixel and then integrated. The integrated magnitude and the surface brightness measurements over the area of the aperture, are derived from the flux measurement. The uncertainty is calculated with the expression\footnote{\url{https://www.gnu.org/software/gnuastro/manual/html_node/Magnitude-measurement-error-of-each-detection.html}}   
\(M_{error} = 2.5\, /\, S/N\, \ln(10)\)
; as the aperture area increases, the signal-to-noise ratio (S/N) also increases, so the magnitude uncertainty decreases. This is different from the flux uncertainty in each pixel (which increases with the square root of the area). However, signal increases linearly with area, so overall, the S/N increases as the area grows larger.

The colour $(g - r)_\mathrm{0}$ was computed for each aperture after correcting the raw magnitudes measured in the two bands for galactic extinction. The Galactic extinction was taken into account using the calibration by \citet{schlafly2011} using the NASA/IPAC Extragalactic Database Extinction Calculator. \footnote{\url{ https://ned.ipac.caltech.edu/extinction_calculator}}. Then the colour $(g - r)_\mathrm{0}$ assigned to a stream is the average of that colour in all the apertures placed on the stream. Our method also allows us to analyse the colour gradient along the streams, although with limited accuracy, as the dispersion in the colour calculation in each aperture can reach 0.1 mag. The colour calculation for the progenitor and the host galaxy is, however, much more precise, due to their higher brightness. 

The detection significance index (DSI), as defined in \textit{Paper1}, is calculated by comparing the signal measurement for a given aperture with the median of a distribution of $N$ random measurements (in our case $N=10000$) for the same-size aperture in the none-detection zone and is given as a factor of the $\sigma$ of that distribution. It is obtained as the output of \textit{--upperlimit-sigma} in \textsc{MakeCatalog} \footnote{\url{https://www.gnu.org/software/gnuastro/manual/html_node/Upper-limit-magnitude-of-each-detection.html}}. Fig. \ref{fig-DSI} (\textit{top}) shows an aperture of effective area  74.41 arcsec$^{2}$ placed on a faint segment of the stream around PGC 597851 and (\textit{bottom}) a histogram of the flux distribution in 10000 apertures of same size, randomly placed over undetected pixels of the image. The DSI for the aperture on the stream is 3.73 $\sigma$ meaning the flux in this faint part of the stream, with a measured surface brightness of 26.91 mag arcsec$^{-2}$ , is 3.73 $\sigma$ above the median of the background noise level for this aperture's footprint.
Columns 9, 10 an 11 of Table \ref{tab:photometry}) show the stream average DSI values for the $r$, $g$ and $z$ bands, respectively. 
Fig. \ref{fig-sbvsdi} shows the trend of decreasing DSI as the streams become fainter for the $r$ band. 
We find similar behaviour for the other bands. 

 \subsection{Stellar stream detectability}
 \label{sec:detectability}

\begin{figure}
\centering
     \includegraphics[width=0.8 \columnwidth]{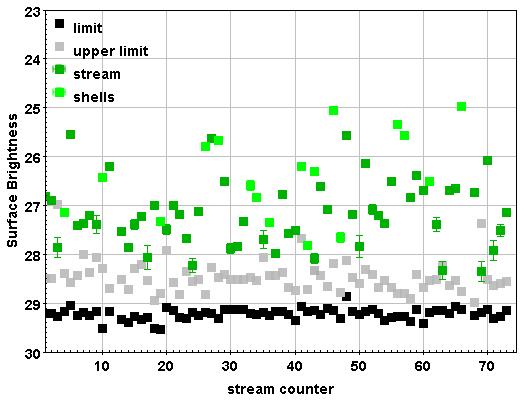}
     \includegraphics[width=0.8 \columnwidth]{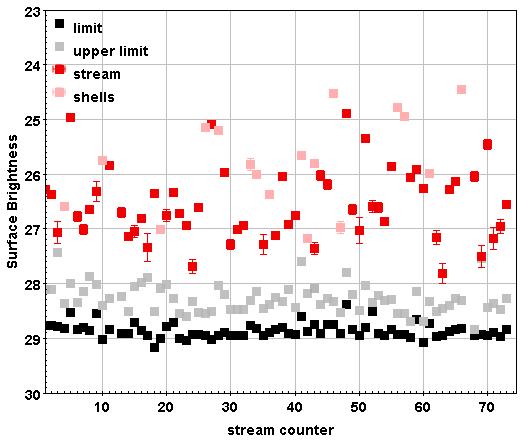}
     \includegraphics[width=0.8 \columnwidth]{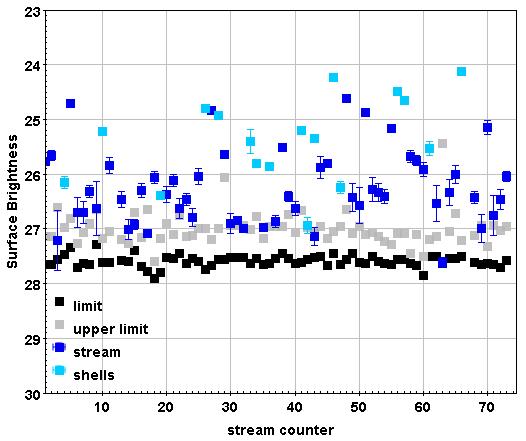}
    \caption{Image surface brightness limit (black) compared with Upper Limit surface brightness (grey) and the faintest surface brightness measured on the streams, excluding shells, for the g-band (green), r-band (red) and z-band (blue), from top to bottom respectively.}
    \label{fig-sbmaxlim}
\end{figure}

\begin{figure}
\centering
     \includegraphics[width=0.8 \columnwidth]{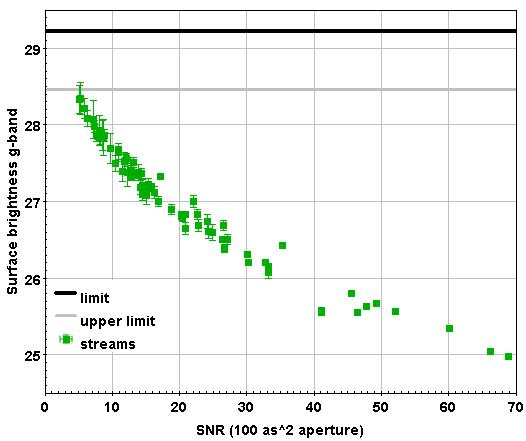}
     \includegraphics[width=0.8 \columnwidth]{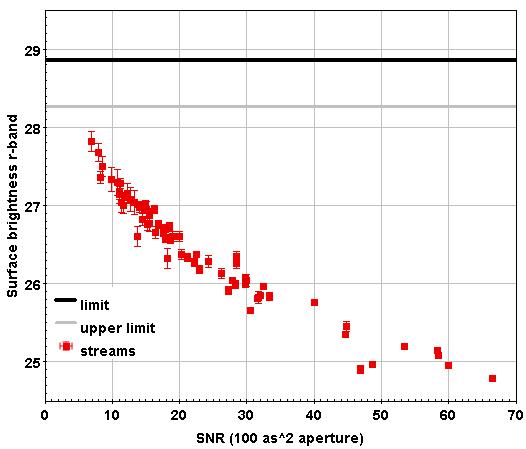}
     \includegraphics[width=0.8 \columnwidth]{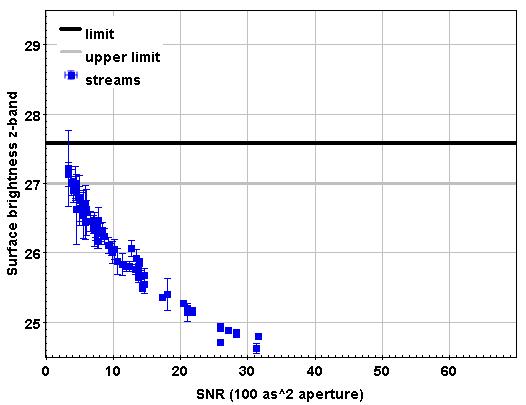}
    \caption{Faintest surface brightness measured on the streams vs SNR, for the g-band (green), r-band (red) and z-band (blue). The solid black line is the average surface brightness limit and the solid grey line is the Upper Limit surface brightness. The \textit{x-axis} represents the ordinal id of the low surface brightness / streams analysed.}
    \label{fig-sbmaxSNR}
\end{figure}

We measured the surface brightness limit of the DES cutout images for the $g$, $r$ and $z$ bands following the definition proposed by \cite{roman2020}, that is, the value corresponding to $+3\sigma$ of the sky background in an area of 100 arcsec$^2$. We report the value as calculated in \textsc{MakeCatalog} \footnote{\url{https://www.gnu.org/software/gnuastro/manual/html_node/Surface-brightness-limit-of-image.html}}, by measuring this limit at pixel level and extrapolating it to the reference area of 100 arcsec$^2$.  

Following this method, the resulting wide-field images reach surface brightness limits  between 28.86 and 29.54 mag arcsec$^{-2}$  for the $g$ band, between 28.37 and 29.17 mag arcsec$^{-2}$  for the $r$ band and between 27.28 and 27.91 mag arcsec$^{-2}$  for the $z$ band. This image depth should be sufficient to be able to detect and characterise faint tidal structures, including new, so far unknown ones.

However, the image depth measured in this way does not fully take into account the correlated noise, likely to be present in the coadded images used here (see background noise histogram in Figure \ref{fig-DSI}). The correlated noise will likely result in an effective increase of the background noise level. The extrapolated pixel-level surface brightness limit is therefore an unrealistically faint estimate of the lowest surface brightness at which streams can be detected robustly. A more realistic measure of the faintest attainable surface brightness can be obtained by measuring the flux in a high number of apertures of 100 arcsec$^{2}$ area placed randomly over the non-detection area of the image and taking the $3\sigma$ value of this distribution ($\sigma$ is the standard deviation of this distribution, that is represented, for PGC 597851, by the histogram of Figure \ref{fig-DSI}). For the rest of this paper, we call the image surface brightness limit measured in this way the Upper Limit Surface Brightness (\textit{ULSB}). The \textit{ULSB} values measured for our DES sample images with stream lie between 26.99 and 28.98 mag arcsec$^{-2}$  in the $g$ band, between 27.44 and 28.84 mag arcsec$^{-2}$  in the $r$ band and between 25.45 and 27.59 mag arcsec$^{-2}$  in the $z$ band. Fig.~\ref{fig-sbmaxlim} shows the surface brightness limits calculated at pixel level and extrapolated to the reference aperture area, the \textit{ULSB} and the surface brightness in the faintest region of the streams measured in this work. 

The difference between the surface brightness limit and the \textit{ULSB} for the $r$ band and the $g$ band is 0.58$\pm$0.22 and 0.75$\pm$0.3 mag arcsec$^{-2}$ , respectively. This difference is mainly due to the presence of structures larger than 1 pixel, 
for example correlated noise. The faintest stream measured in this work is 1.14 mag arcsec$^{-2}$  (and the average stream 2.48 mag arcsec$^{-2}$ ) brighter than the surface brightness limit for the $r$ band. For the $g$ band, the faintest stream is 0.81 mag arcsec$^{-2}$  (the average 2.23 mag arcsec$^{-2}$ ) brighter than the surface brightness limit. The fact that the faintest measured surface brightness is brighter than the \textit{ULSB} is because the uncertainty in the measurement increases as we approach the surface brightness limit and we have set an uncertainty bound of 0.1 mag arcsec$^{-2}$  for the measured surface brightness of the streams. 

This can be seen in Fig.~\ref{fig-sbmaxSNR}, which shows the surface brightness measured in the faintest region of the streams versus the S/N normalised to an aperture of area 100 arcsec$^{2}$ (the surface area used to measure the surface brightness limit). The solid lines indicate the surface brightness limit corresponding to the standard method described above (black) and the \textit{ULSB} (grey), averaged over all the images. 

For example, for the g-band, the average image surface brightness limit is 29.22$\pm$0.11 mag arcsec$^{-2}$  and the average \textit{ULSB} 28.46$\pm$0.32 mag arcsec$^{-2}$  
The figure shows how,
as we go towards fainter streams, the S/N decreases to its minimum value of $~\sim5$ as it 
approaches
the \textit{ULSB}. This is considered to be the true faintness limit for which reliable stream photometry can be measured; below this S/N value, the uncertainty of the measured surface brightness can exceed 0.1 mag arcsec$^{-2}$ , which is the uncertainty limit we have selected for this work.

For the DES images, the difference between the surface brightness limit and the \textit{ULSB} can be explained by the fact that the images are actually a mosaic composed of many exposures, taken under different conditions, on different days, as expected in a large-scale photometric survey
The difference is primarily driven by correlated noise, which is created by mixing the pixels of the individual exposures in the process of rotating, warping and stacking them to compose the final image.


\begin{table*}
\centering
{\small
\caption{Photometry of stellar streams. Columns 1 - 3: host galaxy name and celestial coordinates; column 4: host distance; column 5: host morphological type according to (ref. NASA NED); columns 6: average distance of the stream to the host centre; column 7: width of the stream and column 8: morphological classification, 
A ({\sc Great Circles}), U ({\sc Umbrellas}), GP ({\sc Giant Plumes}), Sp ({\sc Spikes}), PD ({\sc Partially Disrupted Satellite}), Cl ({\sc Clouds}) and NC (Not classified),
see section \ref{sec:morphology}; columns 9 - 11: detection significance index; column 12 indicates whether the stream has been first reported in this work, or in a previous works: (1) \citet{martinez-delgado2023}; (2) \citet{sola2022}; (3) \citet{malin1983}}
\label{tab:photometry}
\setlength{\tabcolsep}{7pt}

\begin{tabular}{lccccccccccc}

Host  & RA &  DEC & D & Type & d & w & m & $\mathrm{DSI}_\mathrm{r}$ & $\mathrm{DSI}_\mathrm{g}$ & $\mathrm{DSI}_\mathrm{z}$ & Reported     \\

  & deg & deg & Mpc &  & kpc & kpc &  & $\sigma$ & $\sigma$ &  $\sigma$ &  \\

\hline\hline

2MASXJ20350262	&	308.7608	&	-44.5274	&	78.34	&		&	35.56	&	2.86	&	GP	&	4.26	&	1.54	&	2.53 & this work	 	\\
ESO 081-008	&	35.4048	&	-64.6103	&	82.41	&	S	&	66.95	&	5.37	&	Cl	&	9.04	&	4.46	&	4.42 & this work	 	\\
ESO 114-001	&	24.4181	&	-60.8653	&	74.82	& SABab	&	29.96	&	3.92	&	PD	&	27.22	&	31.07	&	11.06 & this work	 	\\
ESO147-006	&	340.8275	&	-59.4142	&	71.78	& Sb &	51.24	&	2.86	&	A	&	9.15	&	7.13	&	5.30 & this work	 	\\
ESO 159-013	&	82.4354	&	-53.7556	&	97.72	& Sa &	17.34	&	5.69	&	Cl	&	11.60	&	6.32	&	6.00 & this work	 	\\
ESO 160-002	&	87.813	&	-53.5746	&	61.09	& Sb &	54.09	&	4.23	&	A	&	10.34	&	9.32	&	6.82 & this work	 	\\
ESO 186-063	&	308.5792	&	-54.8467	&	88.72	& Sc &	23.16	&	2.65	&	A	&	9.07	&	4.82	&	2.99 & this work	 	\\
ESO 197-018	&	30.6285	&	-50.9319	&	87.90	& SA0 &	24.49	&	6.79	&	U	&	191.91	&	91.36	&	93.53 & this work	 	\\
ESO 199-012	&	45.8525	&	-50.4881	&	95.94	& S &	27.45	&	4.74	&	GP	&	28.01	&	27.65	&	10.63 & this work	 	\\
ESO 238-024	&	341.2002	&	-49.277	&	99.54	& SABbc	& 47.69	&   6.90    &	NC	&	10.87	&	6.63	&	6.12 & this work	 	\\
ESO 242-007	&	6.0912	&	-45.5073	&	79.07	&	Sc	&	114.11	&	7.81	&	A	&	14.96	&	7.40	&	7.61 & this work	 	\\
ESO 243-006	&	12.9865	&	-43.4776	&	53.46	&	Sb	&	14.89	&	2.30	&	Sp	&	5.78	&	5.80	&	1.57 & this work	 	\\
ESO 285-042	&	309.8261	&	-44.6088	&	95.06	& SABa	&	38.44	&	5.20	&	A+Sp	&	10.23	&	7.82	&	7.69 & this work	 	\\
ESO 287-004	&	319.5263	&	-46.3007	&	59.43	& SBb	&	28.90	&	2.49	&	A/NC	&	4.13	&	4.89	&	1.53 & this work	 	\\
ESO 287-046	&	325.0013	&	-44.0962	&	66.99	& SBc	&	24.36	&	2.97	&	A/NC	&	22.52	&	18.92	&	12.36 & this work	 	\\
ESO 287-051	&	325.4938	&	-44.0917	&	78.34	& Sc	&	21.56	&	2.84	&	U	&	8.60	&	11.03	&	5.41 & this work	 	\\
ESO 340-043	&	308.7697	&	-41.685	&	76.56	& SA0	&	37.18	&	4.35	&	A+NC	&	18.56	&	9.23	&	11.29 & this work	 	\\
ESO 352-073	&	21.3131	&	-33.4088	&	98.17	& Sa	&	28.41	&	7.63	&	U/NC	&	41.09	&	17.76	&	10.04 & this work	 	\\
ESO 476-004	&	20.2808	&	-26.726	&	82.41	& SAB0	&	50.92	&	6.61	&	A	&	17.05	&	7.16	&	6.99 & this work	 	\\
ESO 476-008	&	21.641	&	-23.2271	&	77.98	& SABc	&	45.37	&	5.50	&	U/A	&	6.93	&	5.99	&	5.09 & this work	 	\\
ESO 482-002	&	52.91	&	-25.0088	&	91.20	&	S	&	23.07	&	4.03	&	U/A	&	17.83	&	13.34	&	6.71 & this work	 	\\
ESO 483-012	&	62.5937	&	-23.6173	&	58.08	& S0/a	&	18.67	&	2.34	&	U	&	52.02	&	36.63	&	21.36 & this work	 	\\
ESO 485-011	&	71.637	&	-26.4112	&	76.56	& Irr	&	19.60	&	3.39	&	GP/A	&	55.85	&	32.03	&	23.17 & this work	 	\\
ESO 549-023	&	57.2412	&	-22.1318	&	58.08	& SBa	&	21.00	&	3.79	&	Cl/NC	&	23.03	&	17.21	&	4.18 & this work	 	\\
IC 1657-1	&	18.5288	&	-32.6521	&	48.75	& SBbc	&	52.04	&	4.34	&	GP/NC	&	14.51	&	9.03	&	4.07 & this work	 \\
IC 1657-2	&	    	&               &           &   	&	52.04	&	4.34	&	GP/NC	&	14.51	&	9.03	&	4.07    & this work	\\
IC 1657-3	&	 	    &               &	     	&      	&	52.04	&	4.34	&	GP/NC	&	14.51	&	9.03	&	4.07    & this work	     \\
IC 1816	&	37.9619	&	-36.6734	&	70.15	& SBab	&	40.65	&	2.48	&	A/NC	&	12.54	&	7.68	&	6.23 & this work	 	\\
IC 1833	&	40.4111	&	-28.1722	&	68.87	& SAB0	&	35.25	&	6.27	&	U	&	31.57	&	26.04	&	13.78 & (3)	 	\\
IC 1904	&	48.7532	&	-30.708	&	62.81	& SAB0	&	36.45	&	2.33	&	GP/PD	&	5.37	&	3.29	&	0.95 & (3)	 	\\
IC 2060	&	64.4722	&	-56.6163	&	92.47	& S0	&	47.88	&	8.66	&	U	&	26.29	&	20.75	&	14.16 & this work	 	\\
NGC 1121	&	42.663	&	-1.7342	&	35.81	& S0	&	35.39	&	4.08	&	A	&	18.40	&	8.92	&	6.04 & (2)	 	\\
NGC 1136	&	42.7241	&	-54.9766	&	77.27	& SBa	&	35.21	&	6.33	&	U/A	&	35.87	&	16.73	&	16.87 & this work	 	\\
NGC 1578	&	65.9437	&	-51.5997	&	85.51	& SAa	&	55.75	&	7.44	&	U+GP	&	49.52	&	48.73	&	18.71 & this work	 	\\
NGC 7400	&	343.5864	&	-45.348	&	38.55	& Sbc	&	49.00	&	3.22	&	PD/NC	&	18.36	&	8.58	&	6.81 & this work	 	\\
NGC 7506	&	347.9205	&	-2.161	&	56.49	& SB0	&	19.38	&	3.70	&	U	&	43.23	&	20.60	&	24.43 & this work	 	\\
NGC 823	&	31.7797	&	-25.4439	&	61.09	& SA0	&	58.56	&	5.89	&	U	&	35.01	&	14.03	&	7.29 & this work	 	\\
NGC 922	&	36.2666	&	-24.7905	&	41.69	& SBcd	&	24.54	&	4.37	&	U	&	35.09	&	33.37	&	20.62 & (1)	 	\\
NGC 951	&	37.237	&	-22.3496	&	86.30	& SBab	&	27.38	&	3.16	&	PD	&	28.41	&	20.81	&	11.16 & this work	 	\\
PGC 0122	&	0.3796	&	-40.8196	&	98.17	& S		&	13.24	&	2.83	&	U	&	29.56	&	19.86	&	10.67 & this work	 	\\
PGC 069613	&	341.0415 &	-57.9392 &	63.97	& SA0	&	30.66	&	4.22	&	U	&	138.20	&	97.53	&	46.29 & this work	 	\\
PGC 127531	&	48.731	&	-62.9874	&	81.28	& S		&	36.37	&	3.98	&	A/NC	&	11.04	&	8.15	&	4.40 & this work	 	\\
PGC 127984	&	49.789	&	-61.5145	&	89.13	& S0		&	240.46	&	9.36	&	PD	&	6.59	&	4.08	&	4.77 & this work	 	\\
PGC 1280605	&	11.0258	&	5.2592	&	88.72	&		&	8.97	&	3.52	&	A	&	10.99	&	9.11	&	5.89 & this work	 	\\
PGC 128506	&	19.2737	&	-55.9591	&	66.37	& S		&	9.37	&	3.84	&	NC	&	10.36	&	9.06	&	4.96 & this work	 	\\
PGC 128520	&	25.7061	&	-54.5238	&	89.95	& S0	&	47.72	&	6.27	&	U	&	12.45	&	11.74	&	8.52 & this work	 	\\
PGC 128532	&	34.7909	&	-53.3704	&	87.50	& S0	&	21.42	&	5.52	&	NC	&	35.60	&	25.33	&	26.47 & this work	 	\\
PGC 131085	&	74.5152	&	-39.3889	&	86.70	& S		&	12.27	&	3.45	&	U+A	&	23.95	&	14.35	&	11.28 & this work	 	\\
PGC 131331	&	74.5152	&	-39.3889	&	86.70	& S		&	12.27	&	3.45	&	NC	&	23.95	&	14.35	&	11.28 & this work	 	\\
PGC 131565	&	316.9826 &	-40.7353	&	93.33	& E	    &	11.25	&	4.69	&	U	&	20.38	&	18.44	&	9.70 & this work	 	\\
PGC 132026	&	29.2572	&	-34.1937	&	85.11	& S		&	9.71	&	3.38	&	U	&	28.23	&	19.49	&	12.33 & this work	 	\\
PGC 134111	&	62.8904	&	-27.2743	&	71.12	&	Irr	&	15.66	&	2.83	&	U	&	18.29	&	13.80	&	8.25 & this work	 	\\
PGC 135102	&	58.7478	&	-20.11	&	99.54	& S		&	10.81	&	3.89	&	A	&	8.79	&	7.22	&	3.83 & this work	 	\\
PGC 15602	&	69.0135	&	-23.7295	&	98.17	& S		&	42.40	&	6.57	&	A/NC	&	6.74	&	2.80	&	0.35 & this work	 	\\
PGC 199568	&	44.641	&	-33.8358	&	66.99	&		&	7.56	&	3.08	&	A/U	&	14.85	&	14.36	&	4.34 & this work	 	\\
PGC 3081024	&	84.0906	&	-45.1766	&	97.27	&		&	12.29	&	2.63	&	U	&	67.55	&	53.11	&	26.62 & this work	 	\\
PGC 452979	&	60.4779	&	-51.7533	&	76.56	&		&	9.22	&	1.52	&	NC	&	12.64	&	8.80	&	2.43 & this work	 	\\
PGC 597851-1	&	85.8891	&	-39.4715	&	63.39	&		&	25.87	&	3.68	&	A	&	4.51	&	1.97	&	3.40 & this work	 	\\
PGC 597851-2	&	      	&	         	&	     	&		&	14.84	&	3.34	&	U	&	16.64	&	7.12	&	13.13 & this work	 	\\
PGC 768110	&	67.8642	&	-26.4139	&	71.12	&		&	7.25	&	1.69	&	U	&	25.79	&	15.70	& 10.07 & this work	 	\\
PGC 7743	&	30.5553	&	-6.0789	&	70.15	&		&	11.69	&	2.35	&	PD	&	46.14	&	35.58	&	20.47 & this work	 	\\
PGC 7995	&	31.4528	&	-5.2947	&	71.78	&		&	35.73	&	4.04	&	A	&	30.01	&	19.91	&	11.75 & this work	 	\\
PGC 9063	&	35.7654	&	-1.7489	&	83.95	&		&	12.48	&	3,43	&	PD	&	16.88	&	9.19	&	7.49 & this work	 	\\

\hline

\end{tabular}
}
\end{table*}


\begin{table*}
\centering
{\small
\caption{ Photometry of stellar streams. Column 1 gives the name of the host galaxy. Columns 2 to 4 show the surface brightness in the $g$, in the $r$ and the in the $z$ bands. Columns 5 to 7 show $(g - r)_\mathrm{0}$, $(g - z)_\mathrm{0}$ and $(r - z)_\mathrm{0}$ colours of the streams, averaged over all the apertures placed on the stream}.
\label{tab:photometry2}
\setlength{\tabcolsep}{6pt}

\begin{tabular}{lcccccc}

Host   & $\langle \mu_{r}\rangle_\textrm{stream}$ & $\langle\mu_{g}\rangle_\mathrm{stream}$ & $\langle \mu_{z}\rangle_\textrm{stream}$ & $\langle (g - r)_\mathrm{0} \rangle_\mathrm{stream}$ & $\langle (g - z)_\mathrm{0} \rangle_\mathrm{stream}$ & $\langle (r- z)_\mathrm{0} \rangle_\mathrm{stream}$     \\

  & [mag arcsec$^{-2}$] & [mag arcsec$^{-2}$] & [mag arcsec$^{-2}$] & [mag] & [mag] & [mag]   \\

\hline\hline

2MASXJ20350262		&	26.73	$\pm$	0.04	&	27.44	$\pm$	0.06	&	26.55	$\pm$	0.04	&	0.67	$\pm$	0.07	&	0.81	$\pm$	0.07	&	0.14	$\pm$	0.06	\\
ESO 081-008		&	27.13	$\pm$	0.06	&	27.88	$\pm$	0.09	&	26.63	$\pm$	0.05	&	0.72	$\pm$	0.11	&	1.19	$\pm$	0.10	&	0.48	$\pm$	0.08	\\
ESO114-001		&	25.52	$\pm$	0.05	&	25.88	$\pm$	0.04	&	25.36	$\pm$	0.11	&	0.33	$\pm$	0.06	&	0.46	$\pm$	0.11	&	0.13	$\pm$	0.11	\\
ESO 147-006		&	26.64	$\pm$	0.06	&	27.41	$\pm$	0.09	&	26.18	$\pm$	0.05	&	0.75	$\pm$	0.10	&	1.18	$\pm$	0.10	&	0.43	$\pm$	0.08	\\
ESO 159-013		&	26.53	$\pm$	0.07	&	27.00	$\pm$	0.07	&	26.19	$\pm$	0.14	&	0.41	$\pm$	0.10	&	0.70	$\pm$	0.16	&	0.29	$\pm$	0.16	\\
ESO 160-002		&	26.49	$\pm$	0.05	&	27.02	$\pm$	0.05	&	25.98	$\pm$	0.09	&	0.41	$\pm$	0.07	&	0.83	$\pm$	0.10	&	0.41	$\pm$	0.10	\\
ESO 186-063		&	26.22	$\pm$	0.05	&	27.10	$\pm$	0.06	&	26.03	$\pm$	0.04	&	0.83	$\pm$	0.08	&	0.97	$\pm$	0.07	&	0.14	$\pm$	0.06	\\
ESO 197-018		&	24.13	$\pm$	0.01	&	24.86	$\pm$	0.01	&	23.60	$\pm$	0.01	&	0.71	$\pm$	0.01	&	1.22	$\pm$	0.02	&	0.51	$\pm$	0.01	\\
ESO 199-012		&	25.58	$\pm$	0.04	&	25.98	$\pm$	0.04	&	25.40	$\pm$	0.09	&	0.38	$\pm$	0.05	&	0.54	$\pm$	0.10	&	0.16	$\pm$	0.10	\\
ESO 238-024		&	27.35	$\pm$	0.06	&	26.51	$\pm$	0.04	&	26.23	$\pm$	0.03	&	0.83	$\pm$	0.07	&	1.09	$\pm$	0.06	&	0.27	$\pm$	0.05	\\
ESO 242-007		&	26.96	$\pm$	0.04	&	27.62	$\pm$	0.05	&	26.80	$\pm$	0.04	&	0.64	$\pm$	0.06	&	0.80	$\pm$	0.06	&	0.15	$\pm$	0.05	\\
ESO 243-006		&	26.82	$\pm$	0.08	&	27.04	$\pm$	0.06	&	26.92	$\pm$	0.08	&	0.21	$\pm$	0.09	&	0.10	$\pm$	0.10	&	-0.10	$\pm$	0.11	\\
ESO 285-042		&	26.53	$\pm$	0.06	&	27.04	$\pm$	0.07	&	26.01	$\pm$	0.11	&	0.48	$\pm$	0.09	&	0.97	$\pm$	0.13	&	0.49	$\pm$	0.13	\\
ESO 287-004		&	27.22	$\pm$	0.06	&	27.54	$\pm$	0.06	&	27.09	$\pm$	0.06	&	0.29	$\pm$	0.09	&	0.39	$\pm$	0.08	&	0.10	$\pm$	0.08	\\
ES O287-046		&	26.09	$\pm$	0.05	&	26.68	$\pm$	0.06	&	25.71	$\pm$	0.09	&	0.56	$\pm$	0.07	&	0.92	$\pm$	0.10	&	0.36	$\pm$	0.10	\\
ESO 287-051		&	27.01	$\pm$	0.05	&	27.33	$\pm$	0.04	&	26.39	$\pm$	0.03	&	0.29	$\pm$	0.07	&	0.89	$\pm$	0.05	&	0.61	$\pm$	0.06	\\
ESO 340-043		&	26.14	$\pm$	0.05	&	26.84	$\pm$	0.07	&	26.60	$\pm$	0.08	&	0.65	$\pm$	0.08	&	1.15	$\pm$	0.10	&	0.50	$\pm$	0.09	\\
ESO 352-073		&	26.10	$\pm$	0.04	&	26.83	$\pm$	0.05	&	25.84	$\pm$	0.08	&	0.69	$\pm$	0.06	&	0.91	$\pm$	0.09	&	0.22	$\pm$	0.09	\\
ESO 476-004		&	26.84	$\pm$	0.05	&	27.55	$\pm$	0.07	&	26.36	$\pm$	0.11	&	0.70	$\pm$	0.08	&	1.16	$\pm$	0.12	&	0.47	$\pm$	0.11	\\
ESO 476-008		&	27.43	$\pm$	0.04	&	27.86	$\pm$	0.05	&	26.70	$\pm$	0.02	&	0.41	$\pm$	0.06	&	1.12	$\pm$	0.05	&	0.71	$\pm$	0.04	\\
ESO 482-002		&	26.36	$\pm$	0.06	&	26.80	$\pm$	0.07	&	25.90	$\pm$	0.12	&	0.41	$\pm$	0.09	&	0.86	$\pm$	0.14	&	0.45	$\pm$	0.13	\\
ESO 483-012		&	24.98	$\pm$	0.03	&	25.62	$\pm$	0.03	&	24.60	$\pm$	0.05	&	0.59	$\pm$	0.04	&	0.91	$\pm$	0.06	&	0.33	$\pm$	0.05	\\
ESO 485-011		&	25.04	$\pm$	0.03	&	25.57	$\pm$	0.03	&	24.73	$\pm$	0.05	&	0.49	$\pm$	0.04	&	0.76	$\pm$	0.06	&	0.28	$\pm$	0.05	\\
ESO 549-023		&	25.89	$\pm$	0.03	&	26.42	$\pm$	0.04	&	25.53	$\pm$	0.07	&	0.49	$\pm$	0.05	&	0.79	$\pm$	0.08	&	0.31	$\pm$	0.08	\\
IC 1657-1		&	27.14	$\pm$	0.06	&	27.78	$\pm$	0.08	&	26.67	$\pm$	0.14	&	0.61	$\pm$	0.10	&	1.05	$\pm$	0.16	&	0.44	$\pm$	0.15	\\
IC 1657-2		&	27.01	$\pm$	0.02	&	27.85	$\pm$	0.04	&	26.84	$\pm$	0.03	&	0.81	$\pm$	0.05	&	0.96	$\pm$	0.05	&	0.15	$\pm$	0.03	\\
IC 1657-3		&	26.93	$\pm$	0.04	&	27.32	$\pm$	0.05	&	26.99	$\pm$	0.06	&	0.36	$\pm$	0.07	&	0.28	$\pm$	0.08	&	-0.08	$\pm$	0.07    \\
IC 1816		&	25.59	$\pm$	0.03	&	26.30	$\pm$	0.04	&	25.13	$\pm$	0.02	&	0.68	$\pm$	0.04	&	1.12	$\pm$	0.04	&	0.44	$\pm$	0.03	\\
IC 1833		&	25.67	$\pm$	0.02	&	26.38	$\pm$	0.03	&	25.33	$\pm$	0.05	&	0.70	$\pm$	0.04	&	1.02	$\pm$	0.06	&	0.32	$\pm$	0.05	\\
IC 1904		&	26.95	$\pm$	0.05	&	27.29	$\pm$	0.06	&	26.98	$\pm$	0.06	&	0.33	$\pm$	0.08	&	not rel.			&	not rel.			\\
IC 2060		&	25.97	$\pm$	0.03	&	26.85	$\pm$	0.04	&	25.46	$\pm$	0.05	&	0.86	$\pm$	0.04	&	1.35	$\pm$	0.06	&	0.49	$\pm$	0.05	\\
NGC 1121		&	26.56	$\pm$	0.03	&	27.30	$\pm$	0.05	&	26.40	$\pm$	0.10	&	0.67	$\pm$	0.06	&	0.76	$\pm$	0.11	&	0.09	$\pm$	0.11	\\
NGC 1136		&	25.79	$\pm$	0.03	&	26.49	$\pm$	0.03	&	25.30	$\pm$	0.05	&	0.68	$\pm$	0.04	&	1.15	$\pm$	0.06	&	0.47	$\pm$	0.05	\\
NGC 1578		&	25.86	$\pm$	0.03	&	26.47	$\pm$	0.03	&	25.26	$\pm$	0.05	&	0.60	$\pm$	0.04	&	1.19	$\pm$	0.06	&	0.59	$\pm$	0.05	\\
NGC7400		&	26.75	$\pm$	0.05	&	27.51	$\pm$	0.06	&	26.61	$\pm$	0.14	&	0.74	$\pm$	0.08	&	0.87	$\pm$	0.15	&	0.13	$\pm$	0.14	\\
NGC 7506		&	24.93	$\pm$	0.02	&	25.56	$\pm$	0.02	&	24.49	$\pm$	0.03	&	0.58	$\pm$	0.03	&	0.97	$\pm$	0.04	&	0.39	$\pm$	0.04	\\
NGC 823		&	26.38	$\pm$	0.03	&	27.09	$\pm$	0.05	&	25.89	$\pm$	0.07	&	0.70	$\pm$	0.05	&	1.17	$\pm$	0.08	&	0.48	$\pm$	0.07	\\
NGC 922		&	26.16	$\pm$	0.04	&	26.77	$\pm$	0.05	&	25.89	$\pm$	0.09	&	0.58	$\pm$	0.06	&	0.84	$\pm$	0.10	&	0.25	$\pm$	0.10	\\
NGC 951		&	25.54	$\pm$	0.04	&	26.16	$\pm$	0.05	&	25.18	$\pm$	0.09	&	0.58	$\pm$	0.06	&	0.92	$\pm$	0.10	&	0.34	$\pm$	0.10	\\
PGC 0122		&	24.52	$\pm$	0.04	&	25.05	$\pm$	0.04	&	24.23	$\pm$	0.03	&	0.52	$\pm$	0.04	&	0.80	$\pm$	0.04	&	0.28	$\pm$	0.04	\\
PGC 069613		&	24.45	$\pm$	0.01	&	25.11	$\pm$	0.01	&	24.03	$\pm$	0.02	&	0.64	$\pm$	0.02	&	1.04	$\pm$	0.02	&	0.40	$\pm$	0.02	\\
PGC 127531		&	26.54	$\pm$	0.05	&	27.17	$\pm$	0.06	&	26.34	$\pm$	0.05	&	0.60	$\pm$	0.08	&	0.77	$\pm$	0.08	&	0.17	$\pm$	0.07	\\
PGC 127984		&	27.06	$\pm$	0.05	&	27.83	$\pm$	0.07	&	26.44	$\pm$	0.09	&	0.75	$\pm$	0.09	&	1.34	$\pm$	0.11	&	0.59	$\pm$	0.01	\\
PGC 1280605		&	26.26	$\pm$	0.03	&	26.70	$\pm$	0.03	&	25.92	$\pm$	0.02	&	0.41	$\pm$	0.04	&	0.72	$\pm$	0.04	&	0.31	$\pm$	0.04	\\
PGC 128506		&	26.61	$\pm$	0.06	&	27.21	$\pm$	0.08	&	26.34	$\pm$	0.16	&	0.57	$\pm$	0.10	&	0.82	$\pm$	0.18	&	0.24	$\pm$	0.17	\\
PGC 128520		&	26.87	$\pm$	0.06	&	27.36	$\pm$	0.06	&	26.42	$\pm$	0.12	&	0.47	$\pm$	0.08	&	0.89	$\pm$	0.14	&	0.42	$\pm$	0.14	\\
PGC 128532		&	25.54	$\pm$	0.03	&	26.27	$\pm$	0.03	&	24.91	$\pm$	0.04	&	0.69	$\pm$	0.04	&	1.29	$\pm$	0.05	&	0.60	$\pm$	0.05	\\
PGC 131085		&	26.06	$\pm$	0.05	&	26.78	$\pm$	0.07	&	25.65	$\pm$	0.10	&	0.71	$\pm$	0.08	&	1.10	$\pm$	0.12	&	0.39	$\pm$	0.11	\\
PGC 131331		&	26.06	$\pm$	0.05	&	26.78	$\pm$	0.07	&	25.65	$\pm$	0.10	&	0.71	$\pm$	0.08	&	1.10	$\pm$	0.12	&	0.39	$\pm$	0.11	\\
PGC 131565		&	25.81	$\pm$	0.05	&	26.22	$\pm$	0.05	&	25.57	$\pm$	0.08	&	0.37	$\pm$	0.06	&	0.58	$\pm$	0.09	&	0.21	$\pm$	0.09	\\
PGC 132026		&	25.78	$\pm$	0.05	&	26.27	$\pm$	0.05	&	25.49	$\pm$	0.10	&	0.46	$\pm$	0.07	&	0.74	$\pm$	0.11	&	0.28	$\pm$	0.11	\\
PGC 134111		&	25.89	$\pm$	0.03	&	26.49	$\pm$	0.04	&	25.57	$\pm$	0.03	&	0.56	$\pm$	0.05	&	0.84	$\pm$	0.04	&	0.28	$\pm$	0.04	\\
PGC 135102		&	26.72	$\pm$	0.05	&	27.08	$\pm$	0.06	&	26.43	$\pm$	0.05	&	0.31	$\pm$	0.08	&	0.57	$\pm$	0.08	&	0.25	$\pm$	0.07	\\
PGC 15602		&	27.45	$\pm$	0.06	&	28.05	$\pm$	0.09	&	27.62	$\pm$	0.09	&	0.55	$\pm$	0.11	&	0.33	$\pm$	0.13	&	-0.22	$\pm$	0.11	\\
PGC 199568		&	25.96	$\pm$	0.04	&	26.32	$\pm$	0.04	&	25.90	$\pm$	0.04	&	0.34	$\pm$	0.05	&	0.38	$\pm$	0.05	&	0.04	$\pm$	0.05	\\
PGC 3081024		&	24.45	$\pm$	0.02	&	24.98	$\pm$	0.03	&	24.13	$\pm$	0.02	&	0.49	$\pm$	0.03	&	0.77	$\pm$	0.03	&	0.28	$\pm$	0.03	\\
PGC 452979		&	26.05	$\pm$	0.07	&	26.73	$\pm$	0.09	&	26.43	$\pm$	0.10	&	0.67	$\pm$	0.22	&	0.28	$\pm$	0.13	&	-0.40	$\pm$	0.12	\\
PGC 597851-1		&	26.59	$\pm$	0.05	&	27.23	$\pm$	0.06	&	26.31	$\pm$	0.04	&	0.60	$\pm$	0.08	&	0.83	$\pm$	0.07	&	0.24	$\pm$	0.06	\\
PGC 597851-2		&	25.61	$\pm$	0.03	&	26.23	$\pm$	0.04	&	25.12	$\pm$	0.06	&	0.57	$\pm$	0.05	&	1.02	$\pm$	0.07	&	0.45	$\pm$	0.06	\\
PGC 768110		&	25.46	$\pm$	0.06	&	26.07	$\pm$	0.08	&	25.14	$\pm$	0.13	&	0.58	$\pm$	0.09	&	0.86	$\pm$	0.14	&	0.28	$\pm$	0.13	\\
PGC 7743		&	24.64	$\pm$	0.03	&	25.24	$\pm$	0.04	&	24.20	$\pm$	0.05	&	0.57	$\pm$	0.04	&	0.97	$\pm$	0.06	&	0.41	$\pm$	0.06	\\
PGC 7995		&	25.90	$\pm$	0.04	&	26.50	$\pm$	0.05	&	25.47	$\pm$	0.07	&	0.57	$\pm$	0.06	&	0.98	$\pm$	0.08	&	0.41	$\pm$	0.08	\\
PGC 9063		&	25.91	$\pm$	0.04	&	26.66	$\pm$	0.06	&	25.51	$\pm$	0.03	&	0.73	$\pm$	0.07	&	1.10	$\pm$	0.07	&	0.37	$\pm$	0.05	\\

\hline

\end{tabular}
}
\end{table*}


 \section{DES Stellar Streams Catalogue} 
\label{sec:catalogue}

The catalogue contains the following classes of data:

\begin{itemize}

\item host characteristics, such as the name of the host galaxy, its distance, its K-band and B-band absolute magnitudes, as well as the photometric parameters such as surface brightness and colour.

\item image surface brightness limit in the $r$, $g$ and $z$ bands, as measured in this work. Note: the image surface brightness limit is in itself the 3$\sigma$ value of the sky surface brightness measured in the non-detection zone of the image and extrapolated to an aperture of 100 arcsec$^{2}$; the standard deviation of the sky background value across the image is within 0.1 mag arcsec$^{-2}$  for most of the images, in a few cases being between 0.1 and 0.2 mag arcsec$^{-2}$ . Also the measured \textit{ULSB} values are given (see Section \ref{sec:detectability} for details about this parameter).

\item the stream detection significance index, as defined in Section \ref{sec:dataanalysis} and depicted versus surface brightness for the g-band in \ref{fig-sbvsdi}. 

\item the stream morphology parameters, such as distance to the host, width and length, as well as the morphology class according to the classification given in \citet{martinez-delgado2010} (see Section \ref{sec:morphology} for more details in this classification).

\item stream photometry, such as surface brightness in the the $r$, $g$ and $z$ bands, maximum and averaged over all the apertures placed on the stream. For each stream, the catalogue includes the colours $(g - r)_\mathrm{0}$, $(g - z)_\mathrm{0}$ and $(r - z)_\mathrm{0}$ as an average over the aperture(s), with the respective errors.

\item progenitor characteristics, in those cases where a progenitor is suspected, its coordinates and photometric parameters, such as surface brightness and colours, as well as an estimation of the mass are included.

\end{itemize}

For each stream, the catalogue also indicates whether the stream has been reported for the first time in this work, or in one of the following previous works: (1) \citet{martinez-delgado2023}; (2) \citet{sola2022}; (3) \citet{malin1983}.

A table with all the entries of the catalogue can be found in Appendix \ref{sec:appendix-catalogue}. An extract of the Stellar Streams Legacy Survey catalogue is presented in Table~\ref{tab:photometry} and Table~\ref{tab:photometry2}.

In Appendix \ref{sec:appendix-catalogue} we describe a selection of the detected streams and their hosts, highlighting their characteristics and recognisable morphology, as depicted in Figures \ref{fig-sample1}\textbf{A)}, \ref{fig-sample2}\textbf{B)}, \ref{fig-sample3}\textbf{C)} and \ref{fig-sample4}\textbf{D)}.

\section{Results and Discussion} 
\label{sec:results}

Table \ref{tab:photometry} and Table \ref{tab:photometry2} show the results of our photometric analysis. In the following subsections we will describe the main global characteristics reflected in the results, but without attempting to provide an astrophysical interpretation, which is left for follow-up publications.

 \subsection{Stream Frequency}
 \label{sec:frequency}
 
 We identified tidal streams in 60 galaxy halos from the sample of 689 DES images having a $r$-band surface brightness limit 
 between $28.37$ and $29.17$, mag arcsec$^{-2}$  (see Table \ref{tab:photometry}). This corresponds to  9.1\% $\pm$ 1.1\% of the DES sample galaxies according to a binomial distribution, and implies that, 
 at the 95\% confidence level, 
 the percentage of galaxies in the DES sample that have potentially detectable stellar streams, with the method applied here, is between 6.5\% and 10.9\%. This result matches the result obtained by \citet{morales2018} for 
 streams in a sample of MW-like galaxies in the Local Universe, using images with a surface brightness limit of $\sim$ 28 mag arcsec$^{-2}$ . They reported a total of 28 tidal streams from a sample of 297 galaxies, corresponding to a 9.4\% $\pm$ 1.7\% of galaxies showing evidence of diffuse features that may be linked to satellite accretion events. The result is also consistent with 
 that reported by 
 \citet{miro-carretero2023} where 22 streams were identified 
 in a sample of 181 MW-like galaxies 
 with distances between 25 and 40 Mpc, yielding a detection rate of 12.2\% $\pm$ 2.4 in a similar range of surface brightness for the r-band. While this frequency appears to be slightly higher than we have found in the present work, it is consistent with the expectation that streams are more frequent (and thus more frequently detected at a certain surface brightness limit) in the halos of 
 more massive galaxies, and the MW is on the more massive side of the mass range of galaxies in our sample.
 

\begin{figure*}
\centering
  \includegraphics[width=1.0 \textwidth]{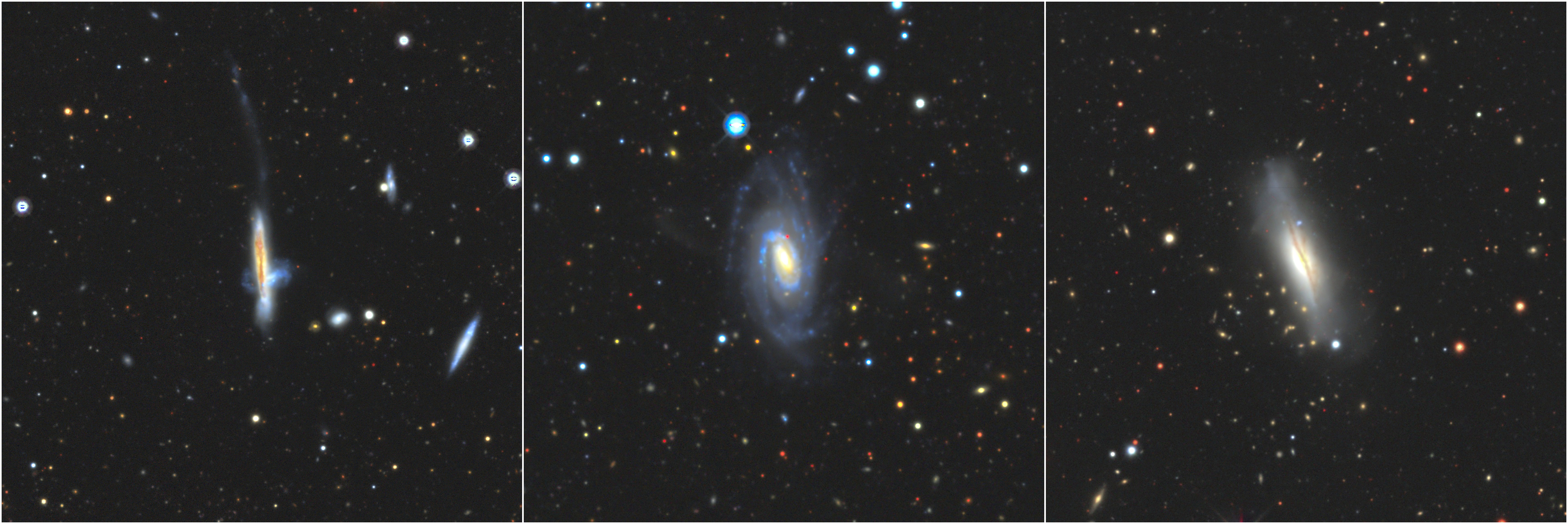}
  \caption{Examples of images with streams as they are obtained from the LS DES catalogue with natural colours; from left to right ESO 199-012, ESO 287-046 and ESO 483-012}
  \label{fig-streamselection}
\end{figure*}

 \subsection{Stream Morphology}
 \label{sec:morphology}


Several classifications of tidal stream morphology have been proposed in the literature \citep{johnston2008,tal2009,martinez-delgado2010,duc2015,giri2023}. In this work, following \textit{Paper1} we adopt a morphological classification based on \citet{johnston2008}, with the extensions proposed in \citet{martinez-delgado2010}: {\sc Great Circles}, hereinafter referred to as circles, are streams that result from satellites along mildly eccentric orbits, with an arc-like shape, sometimes featuring complete loops around the host, but in most cases (in our sample) seen as covering only a small part of a loop; {\sc Umbrellas}, structures often appearing on both sides of the host galaxy, displaying an elongated shaft ending in the form of a shell (sometimes only the shells are visible) resulting from satellites that were on more eccentric, radial orbits; {\sc Giant Plumes}, hereinafter referred to as plumes, structures appearing to shoot out of the host, generally for quite a long distance; {\sc Spikes}, short, elongated jet-like structures emanating from the host galaxy; {\sc Partially Disrupted Satellites}, where the progenitor of the steam can be recognised, and is in the process of being accreted by the host galaxy; {\sc Clouds}, structures presenting a external view with debris spread evenly along only mildly eccentric orbits and {\sc Mixed-type}, displaying smooth large scale structures above and below the plane of the host galaxy, tidal remnants from old accretion events (perhaps > 10 Gyr ago) that have had time to phase mix.

However, this classification implies knowledge of the complete phase-space, which will almost never be available for the far away streams we are studying in this work. Due to the limited depth of our image sample, it is not possible to know how much of the the stream we are actually 
able to detect. Therefore, a precise classification of our findings according to this scheme is generally not possible. 

Notwithstanding this limitation, the apparent morphology of the streams is given in column 8 of Table \ref{tab:photometry}. For approximately half of the streams in our sample we can determine their morphology, according to the classification proposed above, with a reasonable level of confidence: we identify 19 umbrellas (for some of them only the shell part is detectable), seven circles, of which two display complete loops around the host galaxy, three plumes, four partially disrupted satellites, one host galaxy with spikes and one stream of mixed type. For the other half of the stream sample, the morphology cannot be reliably ascertained. For example, in some cases, the structures identified as plumes could be part of a great circle; in other cases several structures with different shapes, visible on the image, could be several streams, or all part of the same stream yielding complex morphology. Some of the detected streams 
cannot be reliably categorised, 
for example faint clouds or short stream segments, as they could be part of a larger structure with a diverse morphology.



The distribution of distance from the host centre to the stream, as an average of the measured distances of the individual apertures placed on the stream, is given in Table \ref{tab:photometry}. More than 90\% of the streams are, on average, within 60~kpc of the host centre, with the rest extending up to 120~kpc. 


The width distribution of the streams, as an average of the diameter of the individual apertures placed on the stream, is given in Table \ref{tab:photometry}. The width of the 
streams ranges between 2 and 8~kpc, with more than half 
having a width less than 4.5~kpc. Our width measurements lie within the range of those reported in \citet{sola2022}, whose Fig.~9 shows the distribution of 
stream width going up to 12~kpc.

 \subsection{Stream Surface Brightness and Colour}
 \label{sec:surfacebrightnessandcolour}

\begin{figure}
\centering
     \includegraphics[width=0.8 \columnwidth]{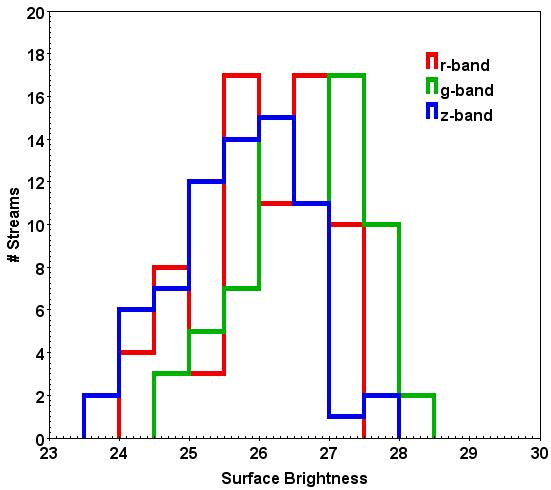}
    \caption{Surface brightness distribution of the analysed stream sample, as listed in Table \ref{tab:photometry2}.}
    \label{fig-histogramsb}
\end{figure}

Table \ref{tab:photometry} shows the average values obtained for each stream for the $g$-band, $r$-band and $z$-band, with the corresponding $1\sigma$ uncertainty. For streams having clearly separated parts or segments, we have measured the photometric parameters separately, and each segment has an entry in the table.
The measured ranges of stream average surface brightness are $24.86 < \mu_{g} < 28.05$ , $24.13 < \mu_{r} < 27.45$ and $23.60 < \mu_{z} < 27.62$ $\mathrm{mag\, arcsec}^{-2}$. 
 
In order to assess the impact of the host galaxy potential (mass) on 
stream characteristics, such as surface brightness and colours, we have compared the results for a subset of the sample, consisting of host galaxies having a MW-like stellar mass, with the overall sample. We have also compared the  $(g - r)_\mathrm{0}$ colour of the streams around these MW-like hosts in the DES sample, with the colour of the streams around MW-like hosts from the SAGA survey \citep{geha2017,mao2021}. To do that, we 
first identified the MW-like hosts within our DES sample, 
using the same criteria as the SAGA survey, namely selecting host galaxies with a K-band absolute magnitude between -23 and -24.6 mag (this magnitude being a proxy of the stellar mass). 32 host galaxies with identified streams belong to this subset ($\sim$ 50 $\%$ of the sample with streams). Then, we compared the colour $(g - r)_\mathrm{0}$ distribution of the MW-like galaxies within the DES sample 
to 
that of the streams identified in the SAGA sample, as reported in \citet{miro-carretero2023}.

\begin{figure}
\centering
     \includegraphics[width=0.8 \columnwidth]{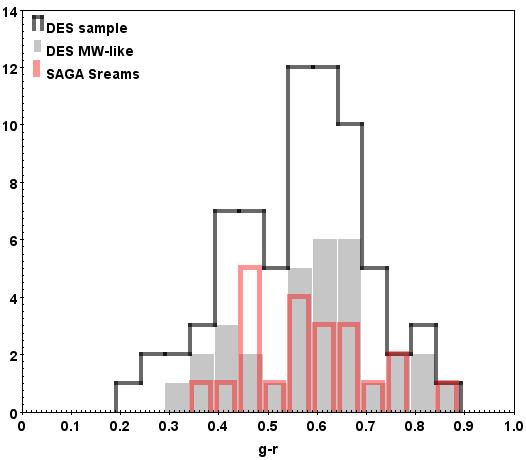}
    \caption{Distribution of colour $(g - r)_\mathrm{0}$ for the analysed stream sample, as listed in Table \ref{tab:photometry2} (black histogram), for the streams around MW-like hosts in the DES sample (filled grey histogram) and for the streams detected in the SAGA sample (red histogram).}
    \label{fig-histogramcolour}
\end{figure}

\begin{figure}
\centering  
     \includegraphics[width=0.8 \columnwidth]{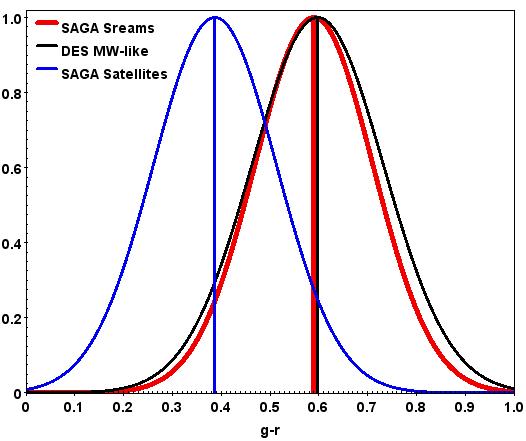}
    \caption{Gaussian fits for colour $(g - r)_\mathrm{0}$ for the analysed stream sample (black contour), for the streams detected in the SAGA sample (red) and for the SAGA satellite galaxies (blue), area-normalised.}
    \label{fig-gaussiancolour}
\end{figure}

Fig.~\ref{fig-histogramcolour} shows the distributions of 
$(g - r)_\mathrm{0}$ colour for our DES stream sample, 
the subset of MW-like host galaxies within this sample, and 
the streams detected in the SAGA sample. As can be seen in this figure, the 
range of $(g - r)_\mathrm{0}$ colour for the 
streams around MW-like host galaxies 
in the DES sample 
is similar to that of
the SAGA sample.
In \citet{miro-carretero2023} it is reported that the hypothesis that the SAGA stream colour distributions follows a Gaussian distribution cannot be rejected with a 99$\%$ confidence according to the result of a contrast of normality test.
Fig.~\ref{fig-gaussiancolour} shows 
a Gaussian fit to the distribution of $(g - r)_\mathrm{0}$ colour for the streams around MW-like hosts within the DES sample, and for the streams around SAGA hosts. 
The average and 
standard deviation 
of the former are 0.60 and 0.13, and 
of the latter 0.59 and 0.12, respectively.
The SAGA and DES MW-like streams therefore have very similar $(g - r)_\mathrm{0}$  colors, and are
significantly redder than 
satellite galaxies from the SAGA survey. This 
strengthens the results reported in \citep{miro-carretero2023}, where such differences between the stream and satellite colour distributions are said to be explained by a combination of selection bias in the SAGA study and physical effects due to tidal stripping.

\begin{figure}
\centering
     \includegraphics[width=0.8 \columnwidth]{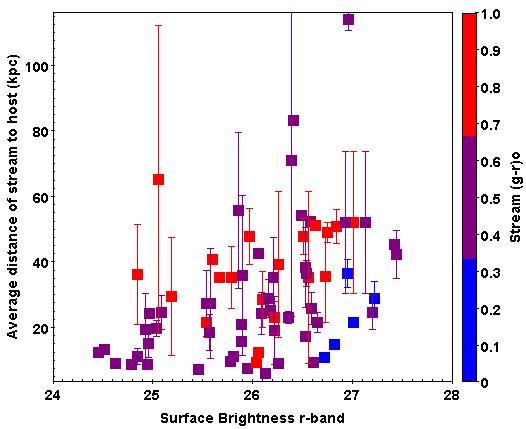}
    \caption{Stream projected distance to the host centre (the points indicate average values and the vertical bars the range of distance along the stream) versus surface brightness in the $r$ band, coloured according to the $\langle (g - r)_\mathrm{0} \rangle$ colour of the stream.}
    \label{fig-distvsSBr}
\end{figure}

Fig. \ref{fig-distvsSBr} shows the projected  distance of the stream from the host centre (the points indicate average values, and the vertical bars the range of distance along the stream) 
plotted against $r$-band surface brightness, 
and colour-coded according to $(g - r)_\mathrm{0}$.  Although we see no strong correlations
between colour, distance to host and surface brightness, there seems to be a trend for the blue streams to be the faintest ($ \langle \mu_{r}\rangle_\textrm{stream} > 26.5 \mathrm{mag\, arcsec}^{-2}$), and for the redder streams to be more than 20 kpc away from the host centre (with a few outliers). We do not attempt to interpret these trends here, because their significance in our present sample is marginal. The distribution shown in Fig. \ref{fig-distvsSBr} demonstrates the potential of the full SSLS dataset for the analysis of such trends.

 \subsection{Host Morphology Type}
 \label{sec:hostmorphology}
 
\begin{figure}
\centering
     \includegraphics[width=0.8 \columnwidth]{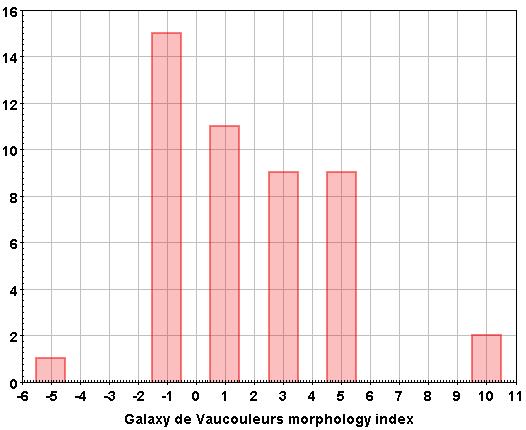}
    \caption{Distribution of streams detected according to the host galaxy morphology given by the galaxy type number: E (-5), S0 (-1), Sa (1), Sb (3), Sc (5) and Irr (10).}  
    \label{fig-histogtype}
\end{figure}

The distribution of streams among the different host galaxy morphology types is shown in Fig.~\ref{fig-histogtype}. We   adopt here the de Vaucouleurs index (deV) to classify the host galaxies by morphology type. We obtain the galaxy classification for each host galaxy from the NED database, and given as a type number for each class: S0 (-1), Sa (1), Sb (3), Sc (5), E (-5) and Irr (10). However, for some host galaxies, the NED database assigns the type with a question mark, indicating that not all reported classifications are in agreement and reflecting the difficulty 
of imposing a discrete classification on a continuous underlying distribution of structural features.

The original sample of 689 galaxies taken as input to our work (see Section \ref{sec:imagesample}) contains $\sim$ 90\% spiral galaxies and $\sim$ 7.5\% S0-type galaxies. However, the percentage of streams detected in spiral galaxies, with respect to the total number of streams detected, is $\sim$ 65\%, and it is $\sim$ 26\% for S0-type galaxies. For the MW-like host galaxies in the DES sample with streams the percentage figures are similar. Thus, S0 galaxies appear much more likely to host detectable streams, compared to other morphological types. One explanation could be that the S0-type galaxies are more massive on average. Indeed, in our sample, the average K-band absolute magnitude of S0s (a proxy for stellar mass) 
is $\sim$ 0.4 mag brighter than 
that of spiral galaxies. 
Among the spiral galaxies, Sa, Sb and Sc galaxies are approximately equally represented in the detected stream sample. We detect streams in only one elliptical galaxy (type E) and in two irregular galaxies (type Irr).

\begin{figure}
\centering
     \includegraphics[width=0.8 \columnwidth]{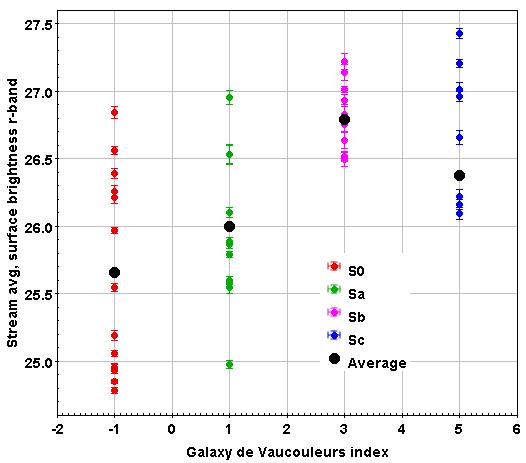}
    \caption{Stream surface brightness versus galaxy morphology, given by the de Vaucouleurs galaxy type number: S0 (-1), Sa (1), Sb (3), Sc (-5).}
    \label{fig-sbvsgtype}
\end{figure}

Fig. \ref{fig-sbvsgtype} shows the distribution of stream surface brightness in our sample versus the host galaxy morphology type, for S0 and Spiral galaxies.
The ranges of 
stream surface brightness for each galaxy type overlap with one another; however, looking at the median surface brightness in each class (black dots), there seems to be a trend 
of streams becoming systematically fainter
from S0 to Sc. In order to verify this hypothesis, we have carried  out an Analysis of Variance\footnote{\url{https://link.springer.com/referenceworkentry/10.1007/978-0-387-32833-1_8}} (ANOVA) for the S0, Sa, Sb and Sc  populations with streams. Under the assumption that 
all these populations have the same variance,
the ANOVA analysis determines that, with 95\% probability, the average of these populations is different. A follow-up Bartlett analysis\footnote{\url{https://link.springer.com/referenceworkentry/10.1007/978-3-642-04898-2_132}} shows however, that the these populations do not have the same variance. We carried out a second ANOVA test, 
now combining the Sb and Sc populations
(because they are not straightforward to distinguish). Following this approach, this second ANOVA test confirmed that the averages of the S0-type, Sa and the (Sb+Sc) populations are different (with a probability of 95\%) and the follow-up Bartlett analysis also confirmed that these populations have the same variance (again with 95\% probability). The results of the ANOVA test are summarised in Appendix \ref{sec:appendix-anova}.

 \subsection{Progenitors}
 \label{sec:progenitors}

\begin{table*}
\centering                          
{\small
\caption{Photometry measurements of the suspected progenitors. Column 1 gives the name of the host galaxy. Column 2 shows the name of the suspected progenitor (if the suspected progenitor does not have a known catalogue name, we give it in this paper the name dwarf-jm followed by an ordinal number n). Column 3 is the projected distance from the centre of the suspected progenitor to the centre of the host galaxy. Columns 4 and 5 show the measured stream surface brightness in the $g$ and in the $r$ bands. Columns 6 and 7 show the $(g - r)_\mathrm{0}$ colour for the progenitor and for the stream (average), respectively. Column 8 shows the range for the estimated minimum progenitor mass. See text for details.}
\label{tab:progenitors1}
\renewcommand{\arraystretch}{1.5}
\begin{tabular}{l c c c c c c c c}       
\hline\hline                 
 Host  & Progenitor & distance & $\mu_{g}$ & $\mu_{r}$ & $(g - r)_\mathrm{0}$  & $\langle (g - r)_\mathrm{0} \rangle_\mathrm{stream}$ & M/M$_{\odot}$\\

 &    & kpc & [mag arcsec$^{-2}$] & [mag arcsec$^{-2}$] & [mag] & [mag] & \\
\hline                        

ESO 114-001	&	PGC 359732	&	28.51	&	22.69	$\pm$	0.01	&	23.21	$\pm$	0.01	&	0.49	$\pm$	0.01	&	0.54	$\pm$	0.04	&	$3.58-6.57$$\times$10$^8$	\\
ESO 417-003	&	dwarf-jm-1	&	31.40	&	24.83	$\pm$	0.02	&	25.52	$\pm$	0.03	&	0.66	$\pm$	0.03	&	0.57	$\pm$	0.06	&	$0.57-1.30$$\times$10$^8$	\\
NGC 951	&	dwarf-jm-2	&	29.72	&	22.70	$\pm$	0.01	&	23.13	$\pm$	0.01	&	0.40	$\pm$	0.01	&	0.58	$\pm$	0.06	&	$3.14-5.06$$\times$10$^8$	\\
PGC 7743	&	dwarf-jm-3	&	15.90	&	23.25	$\pm$	0.01	&	23.77	$\pm$	0.02	&	0.49	$\pm$	0.02	&	0.57	$\pm$	0.04	&	$0.74-2.21$$\times$10$^8$	\\
PGC 9063	&	dwarf-jm-4	&	17.89	&	22.81	$\pm$	0.02	&	23.19	$\pm$	0.02	&	0.35	$\pm$	0.02	&	0.73	$\pm$	0.07	&	$1.04-1.70$$\times$10$^8$	\\
IC 1904	&	LEDA 713873	&	24.22	&	21.67	$\pm$	0.01	&	22.32	$\pm$	0.01	&	0.64	$\pm$	0.01	&	0.58	$\pm$	0.09	&	$2.01-4.50$$\times$10$^8$	\\
NGC 854	&	dwarf-jm-5	&	22.44	&	24.98	$\pm$	0.03	&	25.54	$\pm$	0.04	&	0.54	$\pm$	0.05	&	not reliable	 		&	$4.40-6.74$$\times$10$^7$	\\
NGC 1578	&	dwarf-jm-6	&	155.56	&	24.58	$\pm$	0.02	&	25.25	$\pm$	0.02	&	0.65	$\pm$	0.03	&	not reliable	 		&	$0.58-2.08$$\times$10$^8$	\\
NGC 7400	&	dwarf-jm-7	&	50.06	&	25.44	$\pm$	0.03	&	26.28	$\pm$	0.06	&	0.83	$\pm$	0.06	&	0.74	$\pm$	0.08	&	not reliable	\\

\hline                                   
\end{tabular}
}
\end{table*}


Progenitors were first tentatively identified by visual inspection, tracing apparent overdensities within the stream, and identifying partly disrupted dwarf galaxies in the proximity of the host. We then carried out checks aimed at confirming such progenitors. If the suspected progenitor was a known source, we verified that its distance matched that of the host galaxy; however in most cases the suspected progenitor was not a known source. Then we carried out a photometric analysis of the suspected progenitor by measuring its apparent magnitude, surface brightness and colours. To measure the photometry parameters of a suspected progenitor we followed the same method as for the streams, placing a circular aperture on it and measuring the flux. 
We compared the $(g - r)_\mathrm{0}$ colour of the suspected progenitor with the corresponding colour of the stream and if there was a reasonable match, taking into account the colour dispersion of the stream, the progenitor was confirmed. Note that the colour measurement can have a relatively high uncertainty, and that a colour gradient between the progenitor and the stream can be present. 
The following compact objects were suspected of being progenitors:

\begin{itemize}

\item ESO 114-001: a partially disrupted dwarf galaxy can be seen towards south, at a projected distance of 28.51 kpc from the host galaxy. The object is a known source (PGC 359732) and its distance to the Sun is consistent with it being close to the host galaxy. The stream has a similar $(g - r)_\mathrm{0}$ colour to the suspected progenitor (0.49$\pm$0.01 versus	0.54$\pm$0.04 mag) and thus, the progenitor can be confirmed as likely.
\item ESO 417-003: in this case, the compact object towards east of the host galaxy could be a dwarf at the beginning of being disrupted, as its $(g - r)_\mathrm{0}$ colour is similar to the stream's  average colour (0.66$\pm$0.03	versus	0.57$\pm$0.06 mag) or it could also be a background source. The compact object cannot be confirmed as a progenitor with a high likelihood.
\item IC 1657: three stream segments have been identified around the host on this image. The compact object towards west of the host, although its $(g - r)_\mathrm{0}$ colour matches to one of the stream segments, is a known source, far away in the background (HUbble distance is 	143.66 according to NASA/IPAC Extragalactic Database). The compact object is thus discarded as a progenitor.
\item NGC 951: a rather massive compact object can be seen on the stream, northeast of the host, at a projected distance of 29.72 kpc. The colour difference between the suspected progenitor and the stream is consistent with the colour gradient measured on successive apertures along the stream. The progenitor is confirmed as likely.
\item PGC 7743: the compact object at a projected distance of 15.90 kpc towards south of the host is a suspected progenitor. The $(g - r)_\mathrm{0}$ colour comparison to the stream cannot provide a definite proof since the stream is so close to the host galaxy, the measurement of its colour is likely contaminated, and not reliable. Therefore, the compact object can be assumed to be a progenitor, though not with a high likelihood. 
\item PGC 9063: the compact object at a projected distance of 17.89 kpc, north of the host galaxy could be a progenitor. However, its measured $(g - r)_\mathrm{0}$ colour is much bluer than the average colour of the stream, which in turn is quite similar to the colour of the host. The stream being close to the host galaxy, a contamination of its colour measurement by the host disk cannot be ruled out. Therefore, the compact object can be assumed to be a progenitor, though not with a high likelihood.
\item PGC 127984: the compact object northwest of the host, at a projected distance of 92.76 kpc appears to be within the stream. However, it is a known source (PGC 354833) far away in the background. The oversensity is discarded as a progenitor.
\item IC 1904: a compact object can be seen southeast from the host, between the stream segment and the host. This is a known source (LEDA 713873) at a distance from the Sun consistent with the distance of the host (z $\sim$ 0.0147 vs. z $\sim$ 0.0156). The $(g - r)_\mathrm{0}$ colour of the compact object is very close to the colour of the host and somewhat redder, though within 1 $\sigma$ of the stream colour (0.64$\pm$0.01 versus 0.58$\pm$0.09). The compact object being close to the host galaxy, contamination of its colour measurement by the disk of the host is however likely. Therefore, the progenitor can be confirmed with a high likelihood.
\item NGC 1578: there is a blue compact object towards east inside the host's halo that could be the progenitor, but it is so close to the host that obtaining photometry measurements cannot be done with any accuracy. There is another progenitor candidate on this image; a satellite southwest of the host that is being disrupted, though it is too small for its photometry to be measured reliably, and therefore it cannot be confirmed with high likelihood.
\item NGC 7400: a compact object at 50.06 kpc east of the host galaxy is a potential progenitor. However, its $(g - r)_\mathrm{0}$ colour measures 0.83$\pm$0.06 mag, and , although it is similar to the stream's colour of 0.74$\pm$0.08 mag, it appears too red for it to be confirmed as a likely progenitor.
\item NGC 854: a compact object, towards east, at 22 kpc from the host galaxy centre, is suspected of being the progenitor, but the photometry of the stream could not be measured reliably, and no colour comparison is possible so that the progenitor cannot be confirmed with high likelihood.

\end{itemize}

\begin{figure*}
\centering
 \includegraphics[width=1.0 \textwidth]{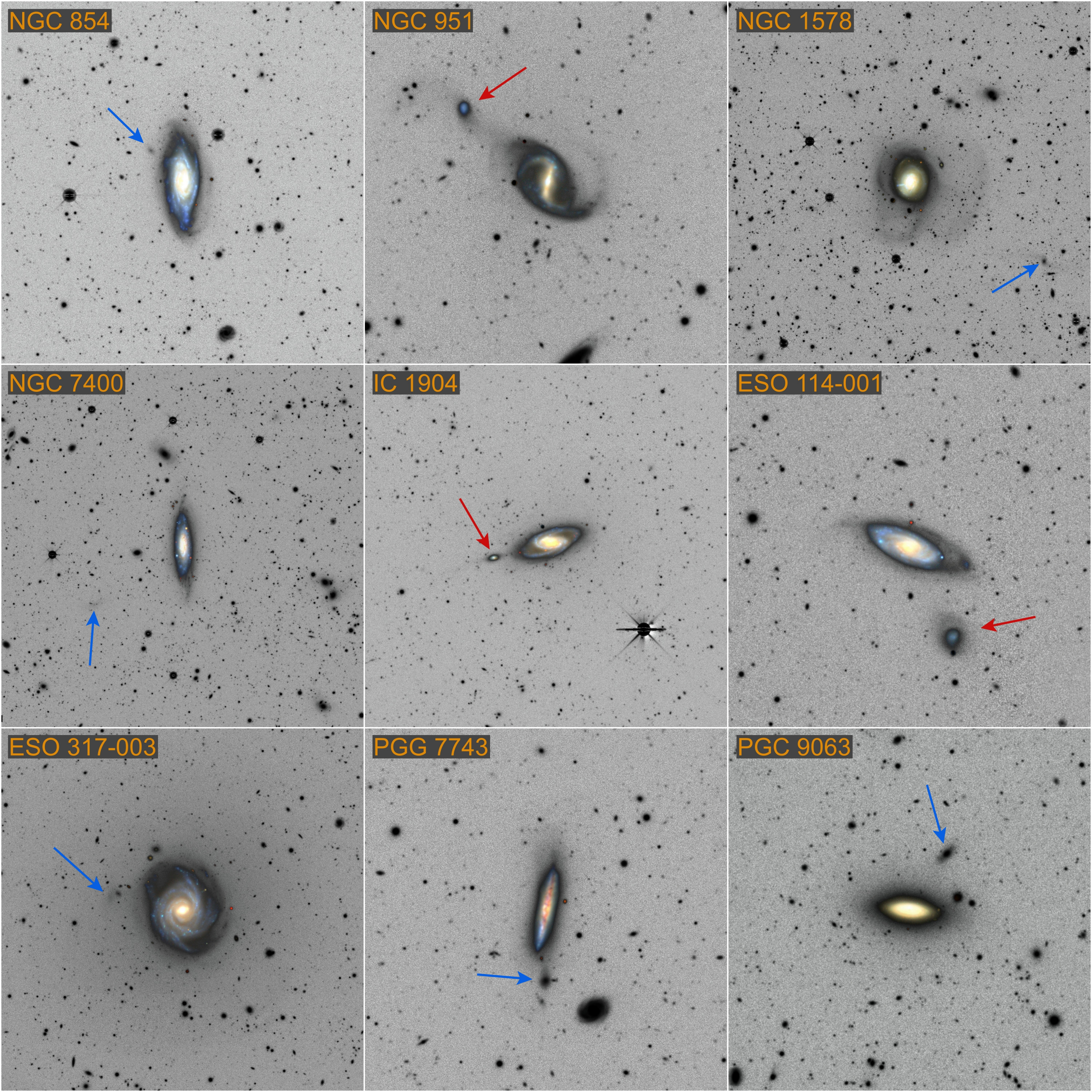}
  \caption{Sample of streams with suspected progenitors. Images with a red arrow indicate the compact object has a high likelihood of being a progenitor. Images with blue arrows indicate the compact object is possibly a progenitor but cannot be confirmed with a high likelihood.}
  \label{fig-progenitors}
\end{figure*}

As a result of the analysis, in nine of the images where we detected streams in our sample, we identified a compact object suspected of being the stream progenitor. Of these, only 3 could be confirmed with a high likelihood. This represents between $\sim$ 5 and 14 \% of the  streams detected. Fig.~\ref{fig-progenitors} shows images with the suspected and confirmed progenitors. Table~\ref{tab:progenitors1} summarises the photometric parameters measured for the suspected stream progenitors. Where the suspected progenitor is not a catalogued object in the NASA/IPAC Extragalactic Database, we give it here the name \textit{dwarf-jm-n, n being a cardinal number}. 

In the cases where the progenitor was confirmed, we have estimated its mass. For those progenitors that are known objects in the NASA/IPAC Extragalactic Database, we estimated their mass from the B-band absolute magnitude, as provided in the Hyperleda Database, using the correlation between B-band magnitude and stellar mass reported in the Spitzer S4G catalogue \citep{munoz-mateos2015}. For the cases where the suspected progenitor was not a catalogued source, the apparent magnitude of the progenitor was measured in a circular aperture  covering as accurately as possible the progenitor. From the apparent magnitudes, the distance modulus of the host (obtained from the Hyperleda database) and the Galactic extinction (obtained from NASA/IPAC Extragalactic Database), we derived the absolute magnitudes for the three bands. Then, we computed the luminosity, using Solar $g$, $r$ and $z$-band absolute magnitudes 
from \citet{willmer2018}.
The uncertainty range of the distance-modulus determines the uncertainty range for the absolute magnitudes.
The results are summarised in Table~\ref{tab:progenitors2}.

We then calculated the mass-to-light ratio from the three colours measured using the correlations between SDSS ugriz colours and SDSS/2MASS M/L ratios given in \citet{bell2003}, and applying the coefficients given in Table A7 of their paper. For each colour, there are three sets of coefficients, one per band, each giving an estimate of the mass. Using the three colours available, we obtain thus nine estimates for the progenitor's mass corresponding to the nine luminosity estimates. As the luminosity estimates have an uncertainty range (inherited from the uncertainty in the distance-modulus) the mass estimation is carried out for the maximum and minimum luminosity values, as well as for the central value. From all the mass estimates obtained, the range of progenitor's mass given in Table~\ref{tab:progenitors1} corresponds to the maximum and minimum mass estimates obtained. The results of the intermediate steps to calculate the distance modulus, absolute magnitudes and mass-to-light ratios are listed in  Table~\ref{tab:progenitors2}.

\begin{figure}
\centering
     \includegraphics[width=0.8 \columnwidth]{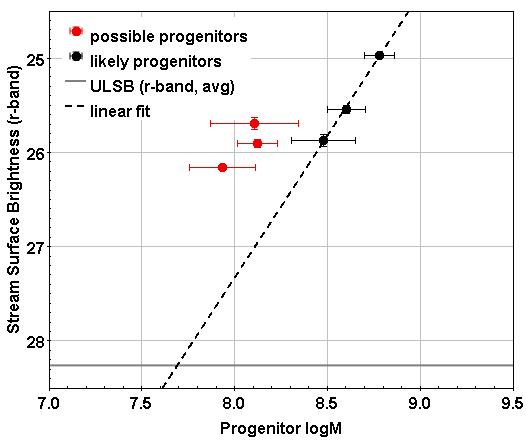}
    \caption{Stream average surface brightness versus progenitor $\log_{10}M_{\star}/\mathrm{M_{\odot}}$). Black dots are the progenitors confirmed with high likelihood; red dots indicate the possible progenitors that cannot be confirmed with high likelihood. The dashed line is the linear fit of the progenitors confirmed with high likelihood. The horizontal grey line indicates the average Upper Limit Surface Brightness for the $r$ band.}
    \label{fig-progenitormass-1}
\end{figure}

\begin{figure}
\centering
     \includegraphics[width=0.8 \columnwidth]{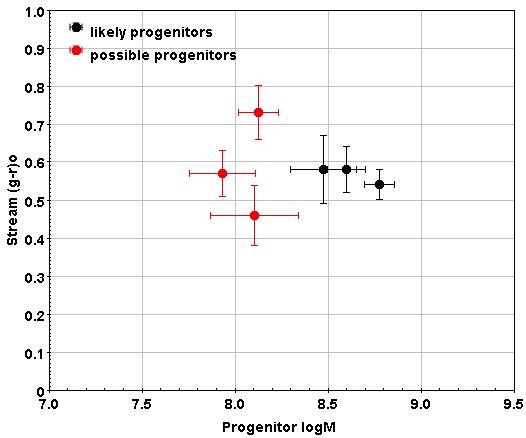}
    \caption{Stream average $(g - r)_\mathrm{0}$ colour versus progenitor log(M$_{*}$/M$_{\odot}$). Black dots are the progenitors confirmed with high likelihood; red dots indicate the possible progenitors that cannot be confirmed with high likelihood.}
    \label{fig-progenitormass-2}
\end{figure}

Figure \ref{fig-progenitormass-1} shows the stream average surface brightness versus the estimated progenitor mass,
$\log_{10} M_\star/\mathrm{M_\odot}$. Black dots correspond to the progenitors confirmed with a high likelihood, while the red dots correspond to the suspected progenitors that cannot be confirmed with a high likelihood. The horizontal grey line indicates the average \textit{ULSB} for the $r$ band. The figure shows that no progenitors have been detected with an estimated 
$\log_{10} M_\star/\mathrm{M_\odot} < 7.75$.
On the other hand, the linear fit of the points corresponding to the confirmed progenitors (black dashed line) crosses the horizontal line representing the \textit{ULSB} at 
$\log_{10} M_\star/\mathrm{M_\odot} < 7.70$,
consistent with the above value. Our interpretation is that progenitors with 
$\log_{10} M_\star/\mathrm{M_\odot} < 7.75$
have a surface brightness below the detectability threshold indicated by the \textit{ULSB}, because they have been partially or completely dissolved by the tidal force. Progenitors above this mass threshold could still be present, but not detected, if occulted by the host galaxy, its bright halo or foreground sources.
Figure \ref{fig-progenitormass-2} shows the stream average $(g - r)_\mathrm{0}$ colour versus the estimated progenitor mass 
($\log_{10} M_\star/M_\odot$). It appears that the streams associated with likely progenitors 
have similar colours, $0.5 < (g - r)_\mathrm{0} < 0.6$.
We see no obvious correlation between stream colour and progenitor mass. Note that the size of the sample is small; a much larger sample would be required to draw any statistically relevant conclusions.

\section{Conclusions and Outlook}
\label{sec:discussion}

In this paper, we present a first-of-its-kind catalogue of 63 stellar streams found in the Local Universe, up to a distance of 100 Mpc. At these distances, and due to the low surface brightness of the streams, very deep images (surface brightness limit > 28 mag arcsec$^{-2}$ ), as well as a dedicated reduction pipeline and image processing tools, have been essential. 
We have taken advantage, on the one hand, of deep images from the Dark Energy Survey (DES) together with the \textit{Legacypipe} reduction pipeline (modified as described Section \ref{sec:imagesample}, and on the other, of the image processing tool GNU Astronomy Utilities (\textsc{Gnuastro}), developed with emphasis on detecting and processing low surface brightness structures. 
The catalogue contains a characterisation of stellar streams regarding morphological aspects, such as distance to the host, shape, and width, as well as photometric measurements, such as surface brightness, colour and detection significance. 
This is the first catalog released within the frame of the extensive \textit{Stellar Streams Legacy Survey} (SSLS), 
which focuses on the census and characterization of minor merger remnants in the Local Universe, as described in \textit{Paper1}.

The main conclusions of the work presented here are as follows:

\begin{itemize}

\item We have, for the first time, characterised photometrically tidal streams up to a distance of 100 Mpc, based on a stellar mass-selected sample of 689 hosts drawn from deep DES images, by measuring their surface brightness in the $r$, $g$ and $z$ bands, together with the respective colours. 

\item 
We have detected 59 
streams (not reported previously) in the Local Universe.

\item We have created a catalogue of 63 tidal streams,
including stream morphology and 
photometric measurements. 
This work extends 
our catalogue of streams in the Local Universe \citep{martinez-delgado2023,miro-carretero2023} to more than 100 entries.

\item  We identified tidal streams in 60 galaxy halos from the sample of 689 DES images (selected on distance, luminosity and isolation criteria, see Section  \ref{sec:imagesample} for details on these criteria). This corresponds to a frequency of  9.1\% $\pm$ 1.1\% of the DES sample galaxies, and implies that with a 95\% confidence level, the percentage of DES sample galaxy halos that have potentially detectable stellar streams, with the method applied here, is between 6.9\% and 11.3\%. This result is consistent with the results obtained in previous work: \citet{morales2018} reported a frequency of 9.4\% $\pm$ 1.7\% of galaxies showing evidence of diffuse features that may be linked to satellite accretion events; \citet{miro-carretero2023} reported a frequency of 12.2\% $\pm$ 2.4 in a similar range of surface brightness for the r-band as the one in our sample. 
Both works refer to detection of streams in samples of MW-like galaxies in the Local Universe. This may explain why the reported frequency appears to be slightly higher than what we have found in the present work, as it is consistent with the 
expectation
that streams are more frequent in the halos of the more massive galaxies, 
combined with the fact that the 
typical galaxy in our DES sample 
has  a lower stellar mass than
the MW ($\log_{10} M_\star/\mathrm{M_\odot}$ = 10$\pm$0.38 for the DES galaxy sample analysed in this work).

\item The colour $(g - r)_\mathrm{0}$ distribution for the streams around MW-like hosts within the DES sample galaxies is in full agreement with 
that reported by \citet{miro-carretero2023} for the streams around SAGA hosts, and is significantly redder than the distribution of $(g - r)_\mathrm{0}$  
for satellite galaxies from the SAGA survey. This strengthens the results reported in that paper, where the differences found between the stream and satellite colour distributions are attributed to a combination of selection effects in the SAGA study and physical effects caused by tidal stripping.

\item The faintest surface brightness measured in streams does not reach the surface brightness limit of the DES images, the latter measured as 3 $\sigma$ of the average brightness at pixel level in the non-detection regions of the image, and then extrapolated to an area of 100 arcsec$^{2}$. 
The faintest stream measured in this work is 1.14 mag arcsec$^{-2}$  
brighter than the surface brightness limit for the $r$ band. For the $g$ band, the faintest stream is 0.81 mag arcsec$^{-2}$  brighter than the surface brightness limit. 
The surface brightness limit calculation above does not fully take into account the flatness of the background, likely not to be perfect for the coadded images used here. Therefore, the depth measured in this way cannot be reached in practice when measuring tidal streams; the actual limiting depth of detectable streams is necessarily brighter.
We have calculated the surface brightness limit in a more realistic way (see Section \ref{sec:detectability} for a detailed explanation). We refer to this more realistic estimate as the Upper Limit Surface Brightness (\textsc{ULSB}). The difference between the surface brightness limit and the \textit{ULSB} is 0.58$\pm$0.22 mag arcsec$^{-2}$  for the $r$ band and 0.75$\pm$0.3 mag arcsec$^{-2}$  the $g$ band, respectively. This difference is primarily driven by correlated noise, which is created by mixing the pixels of the individual exposures in the process of rotating, warping and stacking them to compose the final coadded image. The limiting factor for the measurement of the stream surface brightness to reach the \textit{ULSB} is the measurement uncertainty, which increases 
towards fainter surface brightness. In this work, we set 0.1 mag arcsec$^{-2}$  as the surface brightness maximum measurement uncertainty allowed.


\item For approximately half of the streams in our sample we can determine their morphology type according to the classification proposed by \citet{johnston2008}, and extended by \citet{martinez-delgado2010} with a reasonable level of confidence. We identify (see Section \ref{sec:morphology} for details): 19 umbrellas, seven circles, three plumes, two partially disrupted satellites, or progenitors, one host galaxy with spikes and one stream of mixed type. For the other half of the stream sample, the morphology type cannot be reliably ascertained. We 
remark that the stream morphology is a weak observable, as, on the one hand, due to the depth limitation of our sample images, it is often not possible to detect enough of the stream to determine its shape, and the line of sight of the stream has a significant impact on the morphology perception (see Youdong et al. 2024, in preparation).

\item More than 90\% of the streams are, on average, within 60 kpc
of the host centre, with the 
most distant
parts of streams 
up to 120 kpc. The stream width for 50$\%$ of the streams is 
less than 4 kpc. The maximum width is 
less than 10 kpc.

\item We have investigated a possible correlation between the stream surface brightness and the host galaxy morphology type. Looking at the median stream surface brightness for each morphology type (S0, Sa, Sb, Sc) there is a trend of increasing surface brightness (the stream becoming fainter) as we go from S0 to Sc. 
An Analysis of Variance (ANOVA) has been carried out, in order to verify this hypothesis. Under the assumption that the variance of the populations is the same (confirmed by a Bartlett analysis) the ANOVA test has confirmed that the average for the S0-type, Sa and the (Sb+Sc) populations is different with a probability of 95\% (see Appendix \ref{sec:appendix-anova} for details). 
Note that this is only true after joining the Sb and Sc populations, which is justified by the fact that as a clear differentiation of the two types is not always straight forward, and the purity of these two populations cannot be guaranteed.

\item We identified compact objects as suspected progenitors in 9 of the images with streams in our sample, of which only 3 could be confirmed with a high likelihood. For the others, due to the uncertainties in their distance and 
colour, 
the identification of the progenitor is more speculative.
The number of cases with candidate progenitors represents between $\sim$ 5\% and 14\% of the images with streams and matches earlier estimates of $10\%$ by \citet{miro-carretero2023}. This result 
suggests that most progenitors of the streams detectable at the limiting depth of our survey have been completely dissolved by 
tidal forces, although a number of progenitors may be occulted by the host galaxy, its bright halo or foreground sources. The suspected progenitors have been characterised photometrically and their mass estimated. The mass estimates represent a minimum value, as only the compact object, and not the whole stream has been considered for the calculation. The estimated progenitor masses are in the range $0.5$--$6.6\times10^8\,M_\star/\mathrm{M_{\odot}}$.
No progenitors have been detected with an estimated 
$\log_{10}\,M_\star/\mathrm{M_{\odot}}<7.75$.
Our interpretation is that progenitors less massive than this threshold value have a surface brightness below the detectability threshold indicated by the \textit{ULSB} limit for detection, either from the beginning or because they have been partially or completely dissolved by the tidal force. No correlation between the stream $(g - r)_\mathrm{0}$ colour and the progenitor 
stellar mass
is apparent, perhaps
due to the small 
number of
candidate 
progenitors identified.

\item The photometric analysis of faint, far away streams has been possible thanks to the use of the Gnuastro tool, specifically developed for the detection of faint structures.

\end{itemize}

The catalogue, the product of this work, 
is open to the astronomical community, facilitating further and deeper analysis of the streams 
we have reported, and supporting extension by future studies. 
It constitutes 
a first step in the construction of a more comprehensive catalogue, based on past and future observations. 


Preparation is now underway for low surface brightness science in future ground and space deep surveys, such as ESA's space mission Euclid \citep{euclid2018,hunt2024} and the Vera C. Rubin Observatory \citep{martin2022}. This  includes the characterization of tidal features from mock images, for comparison to larger samples of streams from these surveys. The scrutiny of the images from upcoming large surveys in search of stellar tidal streams will be a daunting task, if performed by human visual inspection
alone.
Therefore, a number of ongoing initiatives are aiming
to develop
machine-learning algorithms 
for this purpose (see for example \citet{ibata2021b,desmons2023,dominguez-sanchez2023}). We believe that our catalogue of streams 
measurements, in combination with the Legacy Survey deep images and the results of processing these images with \textsc{Gnuastro}, could be used as a training set
for the development of these methods.
We anticipate that this 
approach
will be critical for processing the wealth of observations gained from observation projects in the near future, such as those from the Euclid mission, the Vera C. Rubin Observatory, mentioned earlier, The Roman Space Telescope, and others.

The work described here has been carried out 
as part of a 
comprehensive search for stellar streams in the Local Universe, including all the available images within the DESI Legacy Surveys, as 
described
in \textit{Paper1}.
The primary goal of this program is to construct a large, statistically robust dataset suitable for comparison to predictions from simulations of galaxy formation, ultimately to constrain models of baryonic astrophysics and the nature of dark matter on sub-galactic scales.
The work presented here,
using the images from the Dark Energy Survey,
constitutes 
the first significant step in this endeavour. Our work will continue with extensions of the catalogue using the other 
components of the DESI Legacy imaging surveys.
  
\begin{acknowledgements}
        
        We to thank Dustin Lang and John Moustakas for running the modified {\it Legacypipe} code to produce all the images used in this work.
        JMC wants thank the Leiden Observatory for hosting and providing computer infrastructure and facilities for carrying out part of this work, as well as the Universidad Complutense de Madrid for providing computer infrastructure used in this work,
        JMC acknowledges financial support from the Spanish Ministry of Science and Innovation through the project PID2022-138896NB-C55.
        DMD acknowledges the grant CNS2022-136017 funding by MICIU/AEI /10.13039/501100011033 and the European Union NextGenerationEU/PRTR and finantial support from the Severo Ochoa Grant CEX2021-001131-S funded by MCIN/AEI/10.13039/501100011033 and project (PDI2020-114581GB-C21/ AEI / 10.13039/501100011033). 
        MAGF acknowledges financial support from the Spanish Ministry of Science and Innovation through the project PID2022-138896NB-C55.
        APC acknowledges support from a Taiwan Ministry of Education Yushan Fellowship and Taiwan National Science and Technology Council grants 109-2112-M-007-011-MY3 and 112-2112-M-007-017-MY3.
        SRF acknowledge financial support from the Spanish Ministry of Economy and Competitiveness (MINECO) under grant number AYA2016-75808-R, AYA2017-90589-REDT and S2018/NMT-429, and from the CAM-UCM under grant number PR65/19-22462. 
        SRF acknowledges support from a Spanish postdoctoral fellowship, under grant number 2017-T2/TIC-5592. 
        APC is supported by the Taiwan Ministry of Education Yushan Fellowship and Taiwan National Science and Technology Council grant 109-2112-M-007-011-MY3. 
        We acknowledge the usage of the HyperLeda database (http://leda.univ-lyon1.fr).
        The photometry analysis in this work was partly done using GNU Astronomy Utilities (Gnuastro, ascl.net/1801.009) versions $0.17$, $0.18$ and $0.20$. Work on Gnuastro has been funded by the Japanese MEXT scholarship and its Grant-in-Aid for Scientific Research (21244012, 24253003), the European Research Council (ERC) advanced grant 339659-MUSICOS, and from the Spanish Ministry of Economy and Competitiveness (MINECO) under grant number AYA2016-76219-P.
        M.A acknowledges the financial support from the Spanish Ministry of Science and Innovation and the European Union - NextGenerationEU through the Recovery and Resilience Facility project ICTS-MRR-2021-03-CEFCA and the grant PID2021-124918NA-C43.
        This work was coauthored by an employee of Caltech/IPAC under contract No. 80GSFC21R0032.
        
\end{acknowledgements}





\begin{appendix}

\section{Stellar Stream Legacy Survey Catalogue}
\label{sec:appendix-catalogue}

In the following paragraphs we are describing a selection of detected streams, highlighting their measured properties and recognisable morphology in relation to their hosts, as depicted in Figures \ref{fig-sample1}\textbf{A)}, \ref{fig-sample2}\textbf{B)}, \ref{fig-sample3}\textbf{C)} and \ref{fig-sample4}\textbf{D)}. In these figures, colour insets of the central region of the host galaxies have been added to the negative version of the images. For comparison, Figure \ref{fig-streamselection} shows a few examples of images as they are obtained from the LS DES catalogue with natural colours.

\subsection*{2MASXJ20350262-4431375}
\label{2MASXJ20350262-4431375}

 Figure \ref{fig-sample1}\textbf{A)}. SA0-type host galaxy, at a distance of 78.34 Mpc, with a stream of apparent plume morphology. The stream stretches longitudinally from the southeast of the host galaxy with an apparent projected length of $\sim$ 30 kpc, the furthest visible point being at a projected distance of $\sim$ 50 kpc from the host centre, and displays an average width of 2.86 kpc. The average $r$ band surface brightness is 26.73$\pm$0.04 mag arcsec$^{-2}$  and the average $(g - r)_\mathrm{0}$ colour is 0.67$\pm$ 0.07 mag.

\subsection*{ESO 081-008}
\label{sec:ESO081-008}

 Figure \ref{fig-sample1}\textbf{A)}. S-type host galaxy, at a distance of 82.41 Mpc, with a stream in the shape of a cloud at an approximate projected distance of 67 kpc from the host centre in the northeast direction. The  approximate width of the stream is 5.37 kpc. The average $r$ band surface brightness is 27.13$\pm$0.06 mag arcsec$^{-2}$ . Measuring the magnitude for the r and g bands in the circular apertures placed on the stream we obtain an average $(g - r)_\mathrm{0}$ colour of 0.72$\pm$0.10 mag.

\subsection*{ESO 147-006}
\label{sec:ESO147-006}

 Figure \ref{fig-sample1}\textbf{A)}. Sb-type host galaxy, at a distance of 71.78 Mpc, with a stream in the shape of a circle. The stream is at an approximate average projected distance of 51 kpc from the host centre in the southwest direction, and displays an approximate width of 2.86 kpc. The average $r$ band surface brightness is 26.64$\pm$0.06 mag arcsec$^{-2}$ . Measuring the magnitude for the r and g bands in the circular apertures placed on the stream we obtain an average $(g - r)_\mathrm{0}$ colour of 0.75$\pm$0.10 mag.

\subsection*{ESO 159-013}
\label{sec:ESO159-013}

 Figure \ref{fig-sample1}\textbf{A)}. Sa-type host galaxy, at a distance of 97.72 Mpc, with a stream of in the shape of a cloud, on the southeast, at an approximate projected distance of 37 kpc, and a width of 5.69 kpc. The average $r$ band surface brightness  is 26.53$\pm$0.07 mag arcsec$^{-2}$ . Measuring the magnitude for the r and g bands in the circular apertures placed along the stream we obtain an average $(g - r)_\mathrm{0}$ colour of 0.41$\pm$0.10 mag.

\subsection*{ESO 160-002}
\label{sec:ESO160-002}

 Figure \ref{fig-sample1}\textbf{A)}. Sb-type host galaxy, at a distance of 61.09 Mpc, with a stream in the shape of a circle. The stream is at an approximate average projected distance of 54 kpc from the host centre, on the northeast of it, spanning an arc of approximately 20 [deg] on an orbit not centered on the host galaxy. The approximate average width of the stream is 4.22 kpc. The average $r$ band surface brightness is 26.49$\pm$0.05 mag arcsec$^{-2}$ . Measuring the magnitude for the r and g bands in the circular apertures placed on the stream we obtain an average $(g - r)_\mathrm{0}$ colour of 0.41$\pm$0.07 mag.

\subsection*{ESO 186-063}
\label{sec:ESO186-063}

 Figure \ref{fig-sample1}\textbf{A)}. Sc-type host galaxy, edge-on, at a distance of 88.72 Mpc, with a stream of a circular morphology, spanning an arch of approximately half a loop around the host. The apogee of the circle is at a distance of 27.99 kpc from the host centre. This stream, with a projected average stream width of 2.65 kpc, is narrow in comparison to the range of the sample. The $r$ band surface brightness at the apogee is 26.09$\pm$0.10 mag arcsec$^{-2}$ , the average surface brightness in this band being 26.22$\pm$0.05 mag arcsec$^{-2}$ . Measuring the magnitude for the r and g bands in the circular apertures placed along the stream we obtain an average $(g - r)_\mathrm{0}$ colour of 0.83$\pm$ 0.08 mag, placing this stream at the red end of the sample colour spectrum.

\subsection*{ESO 199-012}
\label{sec:ESO199-012}

 Figure \ref{fig-sample1}\textbf{A)}. Sc-type host galaxy, edge-on, at a distance of $95.94$ Mpc, with a bright longitudinal stream reaching out to $\sim$ 45 kpc from the host centre, with a total length of $\sim$ 33 kpc and an average thickness of 4.74 kpc. The surface brightness for the $r$ band shows a gradient from 25.84 $\pm$0.06 at the far side, to 25.42 $\pm$0.03 mag arcsec$^{-2}$  at the nearest point to the host, the average being 25.58 $\pm$0.04 mag arcsec$^{-2}$ . Measuring the magnitude for the $r$ and $g$  bands in the circular apertures placed along the stream we obtain a gradient in the $(g - r)_\mathrm{0}$ colour. Its values are 0.50 $\pm$0.05 at the furthest point from the host and 0.30 $\pm$0.07 at the closest.

\subsection*{ESO 238-024}
\label{sec:ESO238-024}

 Figure \ref{fig-sample1}\textbf{A)}. SABbc-type host galaxy, at a distance of 99.54 Mpc, with a stream of a morphology not clearly identifiable, as only a short segment of it can be detected and characterised. The stream appears to stem from the host galaxy disc at a 45 [deg] inclination in northwest direction and stretching up to 58 kpc from the host centre. The stream displays an average approximate width of 6.90 kpc. The average $r$ band surface brightness is 26.51$\pm$0.04 mag arcsec$^{-2}$ . Measuring the magnitude for the r and g bands in the circular apertures placed on the stream we obtain an average $(g - r)_\mathrm{0}$ colour of 0.83$\pm$0.07 mag, and thus, this stream places itself at the red extreme of the sample's corresponding stream colour range.

\subsection*{ESO 242-007}
\label{sec:ESO242-007}

 Figure \ref{fig-sample1}\textbf{A)}. Sc-type host galaxy, edge-on, at a distance of 79.06 Mpc, with a stream of a circular morphology, with an arch stretching for a projected distance of approximately 15 kpc on a non-concentric orbit at an average projected distance of 110 kpc, in the southeast direction. The average projected width of the circular segment is 7.81 kpc, being thus at the wide end of the sample's stream width range. The $r$ band surface brightness at the apogee is 26.09$\pm$0.10 mag arcsec$^{-2}$ , the average surface brightness in this band being 26.96$\pm$0.04 mag arcsec$^{-2}$ . Measuring the magnitude for the r and g bands in the circular apertures placed along the stream we obtain an average $(g - r)_\mathrm{0}$ colour of 0.64$\pm$ 0.06 mag, placing this stream at the red end of the sample colour spectrum.

\subsection*{ESO 243-006}
\label{sec:ESO243-006}

 Figure \ref{fig-sample1}\textbf{A)}. Sb-type host galaxy, at a distance of 53.46 Mpc, with a stream of a spike morphology. The stream appears to stem from the host galaxy disc in the north direction and stretching up to 16 kpc from the host centre. The stream displays an average approximate width of 2.30 kpc, and thus belongs to the thinner end of the stream sample's width range. The average $r$ band surface brightness is 26.82$\pm$0.08 mag arcsec$^{-2}$ . Measuring the magnitude for the r and g bands in the circular apertures placed on the stream we obtain an average $(g - r)_\mathrm{0}$ colour of 0.20$\pm$0.09 mag, and thus, this stream places itself at the blue extreme of the sample's corresponding stream colour range.

\subsection*{ESO 285-042}
\label{sec:ESO285-042}

 Figure \ref{fig-sample1}\textbf{A)}. SABa-type host galaxy, at a distance of $95.06$ Mpc, with a multi-segment stream in its halo. The segment of the stream going northeast is elongated, reaching out up to $\sim$ 50 kpc from the centre of the host in radial direction with an approximate length of 20 kpc and a thickness of $\sim$ 5 kpc. The other segment of the stream, southwest of the host, is circular, at a distance from the host centre at its perigee of $\sim$ 26 kpc . The for the $r$ band surface brightness for the elongated stream segment shows a gradient from 26.86$\pm$0.07  at the far side, to 26.38$\pm$0.06 mag arcsec$^{-2}$  at the nearest point to the host. The surface brightness measured for the circular stream segment is in average 26.35$\pm$0.06 mag arcsec$^{-2}$ . Measuring the magnitude for the r and g bands in the circular apertures placed along the elongated stream segment we obtain a gradient in the $(g - r)_\mathrm{0}$ colour from 0.69$\pm$0.07 in the proximity of the host galaxy to 0.38$\pm$0.10 at the farthest point from it. The average colour in the elongated part of the stream is 0.49$\pm$0.07 mag, while the average colour in the circular part is 0.46$\pm$0.08 mag. The colour similarity between the elongated segment and the circular segment lets us to believe that the two segments most probably belong to the same stream, rather than being two independent streams.

\subsection*{ESO 287-046}
\label{sec:ESO287-046}

 Figure \ref{fig-sample1}\textbf{A)}. Sc-type host galaxy, at a distance of 66.99 Mpc, with a stream of a circular morphology, spanning what appears to be a loop around the host, perpendicular to its disc plane, although only two discontinuous segments of the loop can be clearly seen. The radius of the suspected circle is approximately 30.69 kpc and its average projected width is 2.97 kpc. The average $r$ band surface brightness  is 26.09$\pm$0.05 mag arcsec$^{-2}$ . Measuring the magnitude for the r and g bands in the circular apertures placed along the stream we obtain an average $(g - r)_\mathrm{0}$ colour of 0.56$\pm$ 0.06 mag.

\subsection*{ESO 287-051}
\label{sec:ESO287-051}

 Figure \ref{fig-sample1}\textbf{B)}. Sc-type host galaxy, face-on, at a distance of $78.34$ Mpc, with an umbrella-type stream of which only one outer shell, disconnected from the disc, can be appreciated clearly. The  shell is at a maximum distance of $\sim$ 23 kpc from the center of the host and covers an arc of $\sim$ 30 degrees. The average surface brightness for the $r$ band is 27.01$\pm$0.05 mag arcsec$^{-2}$ . Measuring the magnitude for the $r$ and $g$ bands in the circular apertures placed on the shell we obtain a value for the $(g - r)_\mathrm{0}$ colour of 0.29$\pm$0.06 mag, placing it at the bluer end of the sample colour spectrum.

\subsection*{ESO340-043}
\label{sec:ESO340-043}

Figure \ref{fig-sample1}\textbf{B)}. SA0-type host galaxy, face-on, at a distance of 74.47 Mpc, with a stream of circular morphology. The stream is visible as an arc on east at $\sim$ with a maximum distance of $\sim$ 46 kpc from the centre of the host galaxy, and as an arc segment on southwest at a maximum distance of $\sim$ 35 kpc from the centre of the host, presumably looping in front of it. Although only discontinuous segments of the stream can be detected, they all have very similar values of the $(g - r)_\mathrm{0}$ colour around 0.72 mag so that it can be determined that they belong to the same stream. The average projected width of the stream is 4.47 kpc and its average $r$ band surface brightness  is 25.19$\pm$0.03 mag arcsec$^{-2}$ .

\subsection*{ESO 352-073}
\label{sec:ESO352-073}

Figure \ref{fig-sample1}\textbf{B)}. Sa-type host galaxy, at a distance of $98.17$ Mpc, with an umbrella-type stream with 2 diffuse outer shells, not quite symmetric, at an angle of $\sim$ 45 deg. with respect to the major axis of the host. The  northeast shell is at a maximum distance of $\sim$ 37 kpc from the center of the host and covers an arc of $\sim$ 30 degrees. The  southwest shell is at a maximum distance of $\sim$ 20 kpc from the center of the host and covers an arc of $\sim$ 30 degrees. Both shells have similar surface brightness, with an average value for the $r$ band of 26.10$\pm$0.05 mag arcsec$^{-2}$ . The average value for the $(g - r)_\mathrm{0}$ colour is somewhat redder for the southwest shell at $\sim$  0.65 mag, while it is at 0.75 mag for the northeast shell.

\subsection*{ESO 476-004}
\label{sec:ESO476-004}

Figure \ref{fig-sample1}\textbf{B)}. SAB0-type host galaxy, at a distance of $82.41$ Mpc, with a visible circular stream segment spanning for $\sim$ 16 kpc at a maximum distance in projection of $\sim$ 56 kpc from the host galaxy centre, and with an average width of 6.61 kpc. The average surface brightness value for the $r$ band is 26.84$\pm$0.05 mag arcsec$^{-2}$  and the average value for the $(g - r)_\mathrm{0}$ colour is 0.70$\pm$0.08 mag.

\subsection*{ESO 476-008}
\label{sec:ESO476-008}

Figure \ref{fig-sample1}\textbf{B)}. SABc-type host galaxy, face-on, at a distance of 77.98 Mpc, with a stream of circular morphology. The stream is visible as an arc at southeast covering $\sim$ 40 [deg] in a circular quasi-concentric orbit around the host's centre with a radius of $\sim$ 45 [kpc]. The average surface brightness value for the $r$ band is 27.43$\pm$0.04 mag arcsec$^{-2}$  and the average value for the $(g - r)_\mathrm{0}$ colour is 0.41$\pm$0.06 mag.



\subsection*{ESO 483-012}
\label{sec:ESO483-012}

Figure \ref{fig-sample1}\textbf{B)}. S0/a-type host galaxy, edge-on, at a distance of $58.07$ Mpc, with an umbrella-type stream displaying possibly several shells, that seem to be in a plane perpendicular to the host disc. Only the outer shell, at a distance of 20 kpc from the host centre, could be characterised photometrically, as the others may be significantly contaminated with the disc. The visible part spans an arc of around 14 kpc longitude at southwest of the host galaxy. The average surface brightness value for the $r$ band is 24.98$\pm$0.03 mag arcsec$^{-2}$  and the average value for the $(g - r)_\mathrm{0}$ colour is 0.59$\pm$0.04 mag.

\subsection*{ESO 485-011}
\label{sec:ESO485-011}

Figure \ref{fig-sample1}\textbf{B)}. Irr-type host galaxy, at a distance of 76.56 Mpc, with a stream in the form of a plume. The stream extends up to a projected distance of $\sim$ 22 kpc from the centre of the host galaxy and has an average width of 3.39 kpc. The average $r$ band surface brightness is 25.04$\pm$0.03 mag arcsec$^{-2}$ , and the average value for the $(g - r)_\mathrm{0}$ colour is 0.49$\pm$0.04 mag.

\subsection*{ESO 544-004}
\label{sec:ESO544-004}

Figure \ref{fig-sample1}\textbf{B)}. Irr-type host galaxy, at a distance of $75$ Mpc, with an LSB structure towards northwest, in the form of an umbrella with a very visible shell, the outer rim of the shell being at a projected distance of 11.01 kpc from the host centre. The photometry measurements on the shell read 24.85$\pm$0.04 mag arcsec$^{-2}$  for the average $r$ band surface brightness, and 0.41$\pm$0.05 mag for the average value for the $(g - r)_\mathrm{0}$ colour. Due to the apparent disturbance of the host galaxy disk and the high brightness of the shell, there is a possibility that this LSB structure is the product of a major merger (\textbf{post merger}). Without ruling out that this structure could be resulting from a minor merger, and in order to ensure the highest possible purity of the stream sample, we do not catalogue this faint structure as a stream without further analysis.

\subsection*{IC 1657}
\label{sec:IC1657}

Figure \ref{fig-sample1}\textbf{B)}. SBbc-type host galaxy, at a distance of $48.75$ Mpc, with three visible streams or stream segments: i) a plume-like extended stream shooting northeast from the host galaxy and stretching up to a distance of $\sim$ 70 kpc from the host centre, ii) a short segment west of the former one, at a distance of $\sim$ 40 kpc from the host centre, that though not visibly connected to it, could be an extension of it and iii) an extended segment southeast of the host galaxy stretching to a distance of 74 kpc from the host galaxy centre, that could be part of the same stream or a separate one.
The plume-like extended stream has an average $r$ band surface brightness of 27.01$\pm$0.02 mag arcsec$^{-2}$ , and an average value for the $(g - r)_\mathrm{0}$ colour is 0.81$\pm$0.05 mag. The short segment towards west has an average $r$ band surface brightness of 26.93$\pm$0.04 mag arcsec$^{-2}$ , and an average value for the $(g - r)_\mathrm{0}$ colour is 0.36$\pm$0.07 mag. A compact object towards west of the host was suspected of being its progenitor, but it turned out to be a catalogued object farther away from the host and discarded as a progenitor (see Section \ref{sec:progenitors}). The third stream towards southeast is long but mostly occulted by foreground stars, and can only be photometrically characterised in a small part, yielding the values of 27.12$\pm$0.05 mag arcsec$^{-2}$  for the surface brightness and 0.61$\pm$0.09 mag for the $(g - r)_\mathrm{0}$ colour. This segment could also possibly be a gas tail (jelly fish). As the colour comparison indicates, the three segments could be separate streams or parts of the same stream that have evolved to having different values for the $(g - r)_\mathrm{0}$ colour.   

\subsection*{IC 1816}
\label{sec:IC1816}

Figure \ref{fig-sample1}\textbf{B)}. SBab-type host galaxy, face-on, at a distance of $70.15$ Mpc, with a umbrella-type stream that extends both on west and east. The west shell is at a distance of $\sim$ 42 kpc from the center of the host and covers an arc of $\sim$ 60 degrees, while the east shell is at $\sim$ 25 kpc from the centre of the host and covers an arc of $\sim$ 120 degrees. The average surface brightness for the $r$ band is 26.18$\pm$0.04 in the west shell and 26.06$\pm$0.05 mag arcsec$^{-2}$  in the east shell. Measuring the magnitude for the $r$ and $g$ bands in the circular apertures placed along both of the shell parts we obtain a measure of the $(g - r)_\mathrm{0}$ colour of 0.52$\pm$0.06 and 0.56$\pm$0.07 mag, respectively. 

\subsection*{IC 1833}
\label{sec:IC1833}

Figure \ref{fig-sample1}\textbf{B)}. SAB0-type host galaxy, face-on, at a distance of $68.87$ Mpc, with an umbrella-type stream of which only one outer shell, disconnected from the disc, can be clearly appreciated. The  shell is at a maximum distance of $\sim$ 36 kpc from the center of the host and covers an arc of $\sim$ 30 degrees. The average surface brightness for the $r$ band is 25.67$\pm$0.02 mag arcsec$^{-2}$ , and the average value for the $(g - r)_\mathrm{0}$ colour is 0.70$\pm$0.04 mag.

\subsection*{IC 1904}
\label{sec:IC1904}

Figure \ref{fig-sample1}\textbf{B)}. SBab-type host galaxy, at a distance of $62.80$ Mpc, with a plume -like stream. Alternatively it could be  a partial disrupted stream from a nearby over-density, initially suspected as the progenitor; however the $(g - r)_\mathrm{0}$ colour being very different between the two, this assumption cannot be confirmed. The plume stretches to a distance of $\sim$ 40 kpc and displays a width of $\sim$ 2.3 kpc. The average surface brightness for the $r$ band is 26.95$\pm$0.05 mag arcsec$^{-2}$ , and the average value for the $(g - r)_\mathrm{0}$ colour is 0.33$\pm$0.08 mag.



\subsection*{NGC 823}
\label{secNGC823}

Figure \ref{fig-sample1}\textbf{C)}. SA0-type host galaxy, at a distance of $61.09$ Mpc, with an umbrella-type stream with two concentric shells on the northwest side and a smaller shell on the east side. The outer northwest shell extends to a projected distance of $\sim$ of 48 kpc, the inner northwest shell to $\sim$ 34 kpc and the east shell is at a projected distance of $\sim$ of 41 kpc. The average surface brightness is different in the different shell layers; for the r-band the measured values are 26.86$\pm$0.05, 25.97$\pm$0.02 and 26.60$\pm$0.04 mag arcsec$^{-2}$  for the northwest outer, northwest inner and east shells respectively. The average values measured for the $(g - r)_\mathrm{0}$ colour are 0.72$\pm$0.08, 0.70$\pm$0.04 and 0.67$\pm$0.06 mag for the three shell layers in the same order as above.

\subsection*{NGC 922}
\label{sec:NGC922}

Figure \ref{fig-sample1}\textbf{C)}. A somewhat irregular spiral host galaxy, face-on, at a distance of $41.69$ Mpc, with an umbrella-type stream showing a shell on one side and the shaft on the other. The shell is at a distance of $\sim$ 20 kpc from the center of the host and covers an arc of $\sim$ 120 degrees, while the shaft goes up to a distance of $\sim$ 29 kpc from the centre of the host. The average surface brightness for the $r$ band is 25.08$\pm$0.00 for the shell and 27.21$\pm$0.03 mag arcsec$^{-2}$  for the shaft. Based on the magnitude for the $r$ and $g$ bands we obtain a measure of the $(g - r)_\mathrm{0}$ colour of 0.53$\pm$0.01 and 0.61$\pm$0.10 mag for the shell and for the shaft, respectively. This stream has been subject to detailed analysis in previous work by \cite{martinez-delgado2023b}.

\subsection*{NGC 951}
\label{sec:NGC951}

Figure \ref{fig-sample1}\textbf{C)}. SBab-type host galaxy, at a distance of $86.30$ Mpc, with extended arms and an bright elongated stream in a northwest radial direction, that extends up to $\sim$ 37 kpc with a total length of $\sim$ 20 kpc and a width of 3.2 kpc. The $r$ band surface brightness shows a gradient from 26.03 $\pm$0.07 at the far side, to 25.32 $\pm$0.05 mag arcsec$^{-2}$  at the nearest point to the host, the average being 25.67$\pm$0.06 mag arcsec$^{-2}$ . Measuring the magnitude for the $r$ and $g$ bands in the circular apertures placed along the stream we obtain a gradient in the $(g - r)_\mathrm{0}$ colour. The average colour of the stream is 0.50$\pm$0.08, although for the most part is 0.54$\pm$0.03. 

An interacting  galaxy, potential precursor is suspected along the stream at $\sim$ 30 kpc from the host centre. Its surface brightness is measured to be 22.71$\pm$0.01 mag arcsec$^{-2}$  and its $(g - r)_\mathrm{0}$ colour 0.40$\pm$0.01 mag.

\subsection*{NGC 1121}
\label{sec:NGC1121}

Figure \ref{fig-sample1}\textbf{C)}. S0-type host galaxy, at a distance of $35.81$ Mpc, wrapped around by a circular stream displaying what appears to be three or more loops around the host. The average stream width is $\sim$ 4 kpc. The $r$ band surface brightness shows a gradient from 26.17$\pm$0.02 at the brightest, to 27.11$\pm$0.05 mag arcsec$^{-2}$  at the faintest, the average being 26.56$\pm$0.03 mag arcsec$^{-2}$ . Measuring the magnitude for the $r$ and $g$ bands in the circular apertures placed along the stream we obtain a gradient in the $(g - r)_\mathrm{0}$ colour. The colour gradient goes from 0.31$\pm$0.07 to 0.83$\pm$0.05 mag, the average being 0.66$\pm$0.07 mag.

\subsection*{NGC 1136}
\label{secNGC1136}

Figure \ref{fig-sample1}\textbf{C)}. SBa-type host galaxy, at a distance of $77.27$ Mpc, with what looks like plumes on opposite sides or an umbrella-type stream whose shells are not completely formed or not visible in their entirety. The northeast plume or shell stretches to a projected distance of $\sim$ 120 kpc from the centre of the host galaxy, while the one at southwest reaches $\sim$ 88 kpc. The average surface brightness for the $r$ band is 25.94$\pm$0.03 mag arcsec$^{-2}$  for the northeast plume and 25.62$\pm$0.02 mag arcsec$^{-2}$  for the southwest one. The average value for the $(g - r)_\mathrm{0}$ colour is 0.68$\pm$0.05 mag and 0.68$\pm$0.04 mag, respectively, reinforcing the assumption that it is an umbrella-type stream with two shells symmetrically located with respect to the host galaxy.

\subsection*{NGC 1578}
\label{sec:NGC1578}

Figure \ref{fig-sample1}\textbf{C)}. SAa-type host galaxy, at a distance of $85.51$ Mpc, with a giant "cloudy" umbrella on the west, and giant plume towards southeast, that could also be a circular stream wrapping around the host on a almost edge-on orbit. The farthest away part of the umbrella is, in projection, at least 80 kpc from the centre of the host galaxy. The measured average stream width is 7.44 kpc. The $r$ band average surface brightness is 25.86$\pm$0.03 mag arcsec$^{-2}$ . Measuring the magnitude for the $r$ and $g$ bands in the circular apertures placed along the stream we obtain the average $(g - r)_\mathrm{0}$ colour 0.60$\pm$0.04 mag. 

\subsection*{NGC 7506}
\label{secNGC7506}

Figure \ref{fig-sample1}\textbf{C)}. SB0-type host galaxy, at a distance of $56.49$ Mpc, with an umbrella-type stream with two shells symmetrically placed on opposite sides of the host galaxy. The shell on the west covers an arc of $\sim$ 30 deg at a distance of $\sim$ 110 kpc from the centre of the host galaxy, and is connected to it with a faint and wide shaft. The shell on the east is less well formed, covering a smaller arc at a distance of $\sim$ 70 kpc. The average surface brightness for the $r$ band is 25.43$\pm$0.03 mag arcsec$^{-2}$  for the west shell and 25.93$\pm$0.03 mag arcsec$^{-2}$  for the east one. The average value for the $(g - r)_\mathrm{0}$ colour is 0.57$\pm$0.04 mag and 0.65$\pm$0.03 mag, respectively.

\subsection*{PGC 0122}
\label{sec:PGC0122}

Figure \ref{fig-sample1}\textbf{C)}. S-type host galaxy, at a distance of $98.17$ Mpc, with an umbrella-type stream of which only one outer shell, can be clearly appreciated. The  shell is at a maximum projected distance of $\sim$ 14 kpc from the center of the host and covers an arc of $\sim$ 50 degrees. The average surface brightness for the $r$ band is 24.51$\pm$0.04 mag arcsec$^{-2}$ , and the average value for the $(g - r)_\mathrm{0}$ colour is 0.52$\pm$0.04 mag.

\subsection*{PGC 7995}
\label{sec:PGC7995}

Figure \ref{fig-sample1}\textbf{C)}. Sbc-type host galaxy, at a distance of $71.78$ Mpc, with a stream of complex morphology, extending in the northwest radial direction from the host up to $\sim$ 60 kpc and displaying a short 'tail' in the opposite direction. The part closer to the host is thick and extended, and at $\sim$ 35 kpc it branches in two elongated segments. The average width of the elongated part of the stream is 4 kpc. The $r$ band surface brightness in this part shows a gradient from 26.28$\pm$0.04 at the farthest end from the host, to 25.59$\pm$0.03 mag arcsec$^{-2}$  at the nearest point to the host. The tail is fainter and difficult to measure as it stretches further away from the host; the measured surface brightness value is 26.45$\pm$0.05 mag arcsec$^{-2}$ . The overall stream average surface brightness in the $r$ band is 25.95$\pm$0.03 mag arcsec$^{-2}$ . Measuring the magnitude for the r and g bands in the circular apertures placed along the stream we obtain the average $(g - r)_\mathrm{0}$ colour. The profile of this colour does not present significant gradients and the average value is 0.57$\pm$ 0.05 mag.

\subsection*{PGC 015602}
\label{sec:PGC015602}

Figure \ref{fig-sample1}\textbf{C)}. S-type host galaxy, at a distance of 98.17 Mpc, with a stream of most likely circular morphology, though the visible part of the stream does not allow a reliable determination of shape of the orbit around the host. The stream describes what appears to be the segment of an arc whose extremes are at $\sim$ 35 and 50 kpc from the host galaxy centre, stretching for $\sim$ 35 kpc, with an average projected width of 6.57 kpc. The average $r$ band surface brightness is 27.45$\pm$0.06 mag arcsec$^{-2}$ , and the average value for the $(g - r)_\mathrm{0}$ colour is 0.55$\pm$0.11 mag.

\subsection*{PGC 069613}
\label{sec:PGC069613}

Figure \ref{fig-sample1}\textbf{C)}. SA0-type host galaxy, at a distance of $63.97$ Mpc, with an umbrella-type stream with two shells placed on opposite sides of the host galaxy. The shell on the west shows what seems to be a shaft and covers an arc of $\sim$ 90 deg at a projected distance of $\sim$ 28 kpc from the centre of the host galaxy. The shaft is however overlapping with what seems to be a host galaxy's arm and cannot be characterised photometrically. The shell on the east seems to be part of a faint halo, concentric with the shell on the east and north, surrounding most of the host galaxy at a projected distance of $\sim$ 34 kpc, however only the brightest part can be photometrically characterised. The average surface brightness for the $r$ band is 24.34$\pm$0.01 mag arcsec$^{-2}$  for the west shell and 24.50$\pm$0.01 mag arcsec$^{-2}$  for the shell on east. The average value for the $(g - r)_\mathrm{0}$ colour is 0.63$\pm$0.02 mag and 0.65$\pm$0.02 mag for the west and east shells, respectively.





\subsection*{PGC 131085}
\label{sec:PGC131085}

Figure \ref{fig-sample1}\textbf{C)}. S0-type host galaxy, at a distance of $79.80$ Mpc, with an umbrella-type stream with two shells placed on opposite sides of the host galaxy. Both shells span an arc of $\sim$ 30 deg at a projected distance of $\sim$ 9.5 kpc from the centre of the host galaxy. The average surface brightness for the $r$ band is 26.26$\pm$0.07 mag arcsec$^{-2}$  for the north shell and 24.95$\pm$0.02 mag arcsec$^{-2}$  for the south shell. The average value for the $(g - r)_\mathrm{0}$ colour is 0.54$\pm$0.03 mag and 0.60$\pm$0.03 mag, respectively.



\subsection*{PGC 132026}
\label{sec:PGC132026}

Figure \ref{fig-sample1}\textbf{D)}. S-type host galaxy, face-on, at a distance of $85.11$ Mpc, with an umbrella-type stream with two shells, an outer and an inner one, placed on opposite sides of the host galaxy. Both shells span an arc of almost 180 deg at a projected distance of $\sim$ 10.18 kpc (outer) and 9.25 kpc (inner). The average surface brightness for the $r$ band is 26.26$\pm$0.07 mag arcsec$^{-2}$  for the outer shell and 25.31$\pm$0.03 mag arcsec$^{-2}$  for the south shell. The average value for the $(g - r)_\mathrm{0}$ colour is 0.41$\pm$0.09 mag for the outer shell and 0.51$\pm$0.05 mag for the inner shell.

\subsection*{PGC 134111}
\label{sec:PGC134111}

Figure \ref{fig-sample1}\textbf{D)}. Irregular spiral host galaxy, at a distance of $71.11$ Mpc, with an umbrella-type stream showing a shell with shaft. The shell is at a distance of $\sim$ 18 kpc from the center of the host and covers an arc of $\sim$ 20 degrees. The average surface brightness for the $r$ band is 25.89$\pm$0.03 for the shell. Based on the magnitude for the $r$ and $g$ bands for the shell, we obtain a measure of the $(g - r)_\mathrm{0}$ colour of 0.56$\pm$0.05 mag.



\subsection*{PGC 597851}
\label{sec:PGC597851}

Figure \ref{fig-sample1}\textbf{D)}. Host galaxy (no classification is provided by the NASA/IPAC Extragalactic Database\footnote{\url{https://ned.ipac.caltech.edu/}}), at a distance of 63.39 Mpc, with a stream of circular morphology, of which only an arc-segment of $\sim$ 90 deg is detected with an apogee (projected distance) of $\sim$ 34 kpc. The average projected width of the stream is 3.68 kpc, its average $r$ band surface brightness is 26.59$\pm$0.05 mag arcsec$^{-2}$  and the average value for the $(g - r)_\mathrm{0}$ colour is 0.60$\pm$0.08 mag.

\begin{table*}[hbt!]
\centering                          
{\small
\caption{Data model of the DES Stream Catalogue}
\label{tab:catalogue}
\renewcommand{\arraystretch}{1.5}
\begin{tabular}{l c l}       
\hline\hline                 
 Parameter  & Units & Description \\

\hline                        

Host-ID	    &	 	&	Host galaxy name	\\
RA  	    & deg 	&	Host Right Ascension		\\
DES 	    & deg 	&	Host Declination		    \\
B-Magnitude & mag 	&	Host absolute magnitude B-band		    \\
K-Magnitude & mag 	&	Host absolute magnitude K-band		    \\
H-SB-r,g,z	    & [mag arcsec$^{-2}$] 	&	Host surface brightness r,g,z-band	         \\
H-SB-r.g.z-err 	& [mag arcsec$^{-2}$] 	&	Host surface brightness r,g,z-band uncertainty \\
H-$(g - r)_\mathrm{0}$    & mag  & Host $(g - r)_\mathrm{0}$ colour \\
H-$(g - r)_\mathrm{0}$-err    & mag  & Host $(g - r)_\mathrm{0}$ colour uncertainty \\
H-$(g - z)_\mathrm{0}$    & mag  & Host $(g - z)_\mathrm{0}$ colour \\
H-$(g - z)_\mathrm{0}$-err    & mag  & Host $(g - z)_\mathrm{0}$ colour uncertainty \\
H-$(r - z)_\mathrm{0}$    & mag  & Host $(r - z)_\mathrm{0}$ colour \\
H-$(r - z)_\mathrm{0}$-err    & mag  & Host $(r - z)_\mathrm{0}$ colour uncertainty \\
H-type      &       & Host galaxy morphology type       \\
H-Distance  &  Mpc  & Host galaxy heliocentric distance \\
SB-lim-r,g,z    & [mag arcsec$^{-2}$] & Image surface brightness limit r,g,z-band \\
UL-SB-r,g,z    & [mag arcsec$^{-2}$] & Image upper limit surface brightness r,g,z-band \\
ST-type    &         & Stream morphological type \\
ST-DIST-H-MAX & arcsec   & Stream maximum distance to host center \\
ST-DIST-H-MIN & arcsec   & Stream minimum distance to host center \\
ST-DIST-H-AVE & arcsec   & Stream average distance to host center \\
ST-Dist-kpc-MAX & kpc   & Stream maximum distance to host center \\
ST-Dist-kpc-MIN & kpc   & Stream maximum distance to host center \\
ST-Dist-kpc-AVE & kpc   & Stream maximum distance to host center \\
ST-WIDTH      & arcsec & Stream average width \\
ST-Width-kpc  & kpc    & Stream average width \\
ST-DI-r,g,z       & $\sigma$ & Stream average Detection Significance Index r-band \\
ST-SB-r,g,z 	    & [mag arcsec$^{-2}$] 	&	Stream average surface brightness r,g,z-band \\
ST-SB-r,g,z-err 	& [mag arcsec$^{-2}$] 	&	Stream average surface brightness r,g,z-band uncertainty \\
ST-SB-max-r,g,z 	    & [mag arcsec$^{-2}$] 	&	Stream maximum surface brightness r,g,z-band \\
ST-SB-max-r,g,z-err 	& [mag arcsec$^{-2}$] 	&	Stream maximum surface brightness r,g,z-band uncertainty \\
ST-$(g - r)_\mathrm{0}$    & mag  & Stream average $(g - r)_\mathrm{0}$ colour \\
ST-$(g - r)_\mathrm{0}$-err    & mag  & stream average $(g - r)_\mathrm{0}$ colour uncertainty \\
ST-$(g - z)_\mathrm{0}$    & mag  & Stream average $(g - z)_\mathrm{0}$ colour \\
ST-$(g - z)_\mathrm{0}$-err    & mag  & Stream average $(g - z)_\mathrm{0}$ colour uncertainty \\
ST-$(r - z)_\mathrm{0}$    & mag  & Stream average $(r - z)_\mathrm{0}$ colour \\
ST-$(r - z)_\mathrm{0}$-err    & mag  & Stream average $(r - z)_\mathrm{0}$ colour uncertainty \\
SNR-min-r,g,z &   & minimum S/N measured in apertures on the stream \\
SNR-min-norm-r,g,z &   & minimum S/N measured, normalised for an aperture of 100 arcsec$^2$ on the stream \\

\hline                                   

\end{tabular}
}
\end{table*}

\newpage

\section{ANOVA for the surface brightness distribution of host morphology types}
\label{sec:appendix-anova}

In Section \ref{sec:hostmorphology}, the assumption is made that there is a correlation between the host galaxy morphology type and the surface brightness, such that the surface brightness increases as the morphology type goes from S0 to Sc. In order to verify this hypothesis, an Analysis of Variance (ANOVA) has been carried out for S0, Sa and Sb+Sc populations (Sb and Sc populations have been merged for this analysis, as a clear differentiation between these populations is not always straight forward, and the purity of these two populations cannot be guaranteed). Table \ref{tab:anova} includes the input variables for the analysis for the three populations.

\begin{table}[hbt!]
\centering                          
{\small
\caption{Analysis of Variance (ANOVA) input variables.}
\label{tab:anova}
\begin{tabular}{lcccc}       
\hline\hline                 
  & S0 & Sa & Sb+c & T \\

\hline                        

T$_{j}$                 & 359.193 &	311.912 & 742.171 & 1413.28	\\
$\Sigma x_{ij}^2$       & 9222.24 &	8110.92 & 19684.4 & 37017.6	\\
n$_{j}$                 & 14      &	12      & 28      & 54	    \\
$\sigma_{ij}^2$          & 0.50304 &	0.31951 & 0.45681 &     	\\

\hline                                   
\end{tabular}
}
\end{table}

where: 

$j$ stands for the population (S0, Sa, Sb+Sc), n$_{j}$ for the number of galaxies in that population, $i$ are the members of that population, T is the total number of observations and,

\[T_{j} = \Sigma_{i=1}^{n_{j}} x_{ij} \]
\[\Sigma x_{ij}^2 = \Sigma_{i=1}^{n_{j}} x_{ij}^2 \]
\[\sigma_{ij}^2 = \frac{1}{n_{j}-1}(\Sigma_{i=1}^{n_{j}} x_{ij}^2 - \frac{T_{j}^2}{n_{j}} )\]
\\
The calculation of the statistical sums for the group factors and error factors is as follows, SSE being the Sum of Square Residuals and SSG the Sum of Square Groups:

\[SSG = \Sigma_{i=1}^{k}\frac{T_{j}^2}{n_{j}} - \frac{T^2}{n} = 7.22993\]
\[SSE = \Sigma_{i=1}^{k} ( \Sigma_{i=1}^{n_{j}} x_{ij}^2 - \frac{T_{j}^2}{n_{j}}) = 22.388\] 
\[\sigma_{MSG}^2 = \frac{SSG}{k-1} = 3.61497\]
(with k=3)
\[\sigma_{MSE}^2 = \frac{SSE}{n-k} = 0.43898\]
(with n=54)
\newline
\\
and with this, the f ratio as used in the Fisher test:

\[f_{sample} = \frac{\sigma_{MSG}^2}{\sigma_{MSE}^2} = 8.23493\]
\\
and since the distribution function of the standard normal distribution  \[F(0.05,2,51) = 3.1788\]
the hypothesis can be confirmed with a 95\% probability as f$_{sample}$ > F(0.05,2,51). However, this result is only valid if the different populations have the same variance. In order to verify this we apply the Bartlett test and obtain:

\[V_{sample} = 0.64857\]
\[C_{sample} = 1.03088\]
\[\chi_{sample}^2 = \frac{V_{sample}}{C_{sample}} = 0.62915\]
\\
on the other hand

\[\chi^2 (0.05;3) = 5.99146\]
\\
As $\chi_{sample}^2$ is lower than the acceptance region limit $\chi^2(0.05;3)$ it can be confirmed that the S0-type, Sa and the (Sb+Sc) populations have the same variance (with a probability of 95\%), and therefore also, the populations S0, Sa and Sb+Sc have different average surface brightness limit (with the same probability).

\section{Characteristics of Identified Progenitors}
\label{sec:appendix-progenitors}

In Section \ref{sec:progenitors} we explain how we estimated the mass of the suspected progenitors. The results of the intermediate steps to calculate the distance modulus, absolute magnitudes and mass-to-light ratios are listed in  Table~\ref{tab:progenitors2}.

\begin{table*}
\centering                          
{\small
\caption{Absolute magnitude and luminosity of suspected progenitors. Column 1 Distance modulus. Columns 2 to 4 absolute magnitude for the $r$, $g$ and $z$ bands. Columns 5 to 7 luminosity for the $r$, $g$ and $z$ bands}
\label{tab:progenitors2}
\renewcommand{\arraystretch}{1.5}
\begin{tabular}{l c c c c c c c}       
\hline\hline                 
 Progenitor  & m-M & M$_{r}$ & M$_{g}$ & M$_{z}$ & L$_{r}$/L$_{\odot}$$_{r}$  & L$_{g}$/L$_{\odot}$$_{g}$ & L$_{z}$/L$_{\odot}$$_{z}$ \\

 & [mag] & [mag] & [mag] & [mag] &   &  & \\
\hline                        

PGC 359732	&	34.37$\pm$0.21	&	-16.60	$\pm$	0.21	&	-16.15	$\pm$	0.21	&	-16.91	$\pm$	0.21	&	$3.16	^{+0.67}_{-0.55}$$\times$10$^8$	&	$3.01	^{+0.64}_{-0.53}$$\times$10$^8$	&	$3.66	^{+0.78}_{-0.64}$$\times$10$^8$	 	\\
dwarf-jm-1	&	34.32$\pm$0.22	&	-14.41	$\pm$	0.22	&	-13.74	$\pm$	0.22	&	-14.83	$\pm$	0.22	&	$4.04	^{+0.91}_{-0.74}$$\times$10$^7$	&	$3.29	^{+0.74}_{-0.60}$$\times$10$^7$	&	$5.39	^{+1.21}_{-0.99}$$\times$10$^7$	 	\\
dwarf-jm-2	&	34.68$\pm$0.18	&	-16.60	$\pm$	0.18	&	-16.17	$\pm$	0.18	&	-16.87	$\pm$	0.18	&	$3.06	^{+0.55}_{-0.47}$$\times$10$^8$	&	$3.08	^{+0.56}_{-0.47}$$\times$10$^8$	&	$3.54	^{+0.64}_{-0.54}$$\times$10$^8$	 	\\
dwarf-jm-3	&	33.97$\pm$0.49	&	-15.00	$\pm$	0.49	&	-14.51	$\pm$	0.49	&	-15.28	$\pm$	0.49	&	$6.99	^{+4.99}_{-2.54}$$\times$10$^7$	&	$6.68	^{+3.81}_{-2.42}$$\times$10$^7$	&	$8.13	^{+4.64}_{-2.95}$$\times$10$^7$		\\
dwarf-jm-4	&	34.62$\pm$0.19	&	-15.50	$\pm$	0.19	&	-15.15	$\pm$	0.19	&	-15.69	$\pm$	0.19	&	$1.11	^{+0.21}_{-0.18}$$\times$10$^8$	&	$1.2	^{+0.23}_{-0.19}$$\times$10$^8$	&	$1.20	^{+0.23}_{-0.19}$$\times$10$^8$		\\
LEDA 713873	&	33.99$\pm$0.25	&	-16.50	$\pm$	0.25	&	-16.25	$\pm$	0.25	&	-16.90	$\pm$	0.25	&	$2.79	^{+0.72}_{-0.57}$$\times$10$^8$	&	$2.33	^{+0.60}_{-0.48}$$\times$10$^8$	&	$3.62	^{+0.94}_{-0.75}$$\times$10$^8$		\\
dwarf-jm-5	&	34.69$\pm$0.18	&	-14.05	$\pm$	0.18	&	-13.51	$\pm$	0.18	&	-14.45	$\pm$	0.18	&	$2.92	^{+0.52}_{-0.45}$$\times$10$^7$	&	$2.65	^{+0.48}_{-0.41}$$\times$10$^7$	&	$3.81	^{+0.69}_{-0.58}$$\times$10$^7$		\\
dwarf-jm-6	&	34.66$\pm$0.19	&	-14.97	$\pm$	0.19	&	-14.32	$\pm$	0.19	&	-15.18	$\pm$	0.19	& 	$6.80	^{+1.30}_{-1.10}$$\times$10$^7$	&	$5.59	^{+1.07}_{-0.90}$$\times$10$^7$	&	$7.43	^{+1.42}_{-1.19}$$\times$10$^7$		\\
dwarf-jm-7	&	32.92$\pm$0.28	&	-11.97	$\pm$	0.28	&	-11.13	$\pm$	0.28	&	-12.62	$\pm$	0.28	&	$4.30	^{+1.27}_{-0.98}$$\times$10$^6$	&	$2.98	^{+0.88}_{-0.68}$$\times$10$^6$	&	$7.05	^{+2.07}_{-1.60}$$\times$10$^6$	 	\\

\hline                                   
\end{tabular}
}
\end{table*}

\end{appendix}

\end{document}